# From Principles to Rules:
# A Regulatory Approach for Frontier AI

Jonas Schuett,[1)✉] Markus Anderljung,[1),2)] Alexis Carlier,[3)]
Leonie Koessler,[1)] Ben Garfinkel[1),4)]

## Abstract

Several jurisdictions are starting to regulate frontier artificial intelligence (AI) systems, i.e. general-purpose AI systems that match or exceed the capabilities present in the most advanced systems. To reduce risks from these systems, regulators may require frontier AI developers to adopt safety measures. The requirements could be formulated as high-level principles (e.g. 'frontier AI systems should be safe and secure') or specific rules (e.g. 'frontier AI systems must be evaluated for dangerous model capabilities following the protocol set forth in…'). These two regulatory approaches, known as 'principle-based' and 'rule-based' regulation, have complementary strengths and weaknesses. While specific rules provide more certainty to developers and are easier to enforce, they can quickly become outdated and lead to a box-ticking attitude. Conversely, while high-level principles provide less certainty and are more costly to enforce, they are more adaptable and more appropriate in situations where the regulator is unsure exactly what behavior would best advance a given regulatory objective. However, rules-based and principles-based regulation are not binary options. Policymakers must choose an appropriate point on the spectrum between them, recognizing that the right level of specificity may vary between requirements and change over time.

Against this background, we recommend that policymakers should initially (1) mandate adherence to high-level principles for safe frontier AI development and deployment (though some requirements could already be formulated as rules), (2) ensure that regulators and third parties closely oversee how developers comply with these principles, and (3) urgently build up regulatory capacity. Over time, the approach should likely become more rule-based. Our recommendations are based on the assumptions that (A) risks from frontier AI systems are poorly understood and rapidly evolving, (B) many safety practices are still nascent, (C) frontier AI developers are best placed to innovate on safety practices, (D) developers' incentives are not aligned with the public interest, (E) regulators have limited expertise and access to information, (F) regulators are better placed to assess the adequacy of company practices than to set detailed rules, and (G) frontier AI developers can handle regulatory uncertainty. Should these assumptions be contested, a more rules-heavy approach might be preferable.

---

[1)] Centre for the Governance of AI, Oxford, UK. [2)] Center for a New American Security, Washington, DC, US. [3)] RAND, Washington, DC, US. [4)] University of Oxford, Oxford, UK.
✉ Contact: jonas.schuett@governance.ai.

# I. Introduction

The US, UK, and EU are starting to regulate frontier artificial intelligence (AI) systems, i.e. highly capable general-purpose AI systems that can perform a wide variety of tasks and match or exceed the capabilities present in the most advanced systems.[1] In the US, a number of frontier AI developers have made voluntary safety commitments,[2] after several hearings in Congress[3] and a meeting in the White House.[4] In late 2023, the Biden administration also issued a landmark Executive Order on Safe, Secure, and Trustworthy AI, imposing reporting requirements on systems more advanced than any that exist today.[5] In the UK, efforts to regulate AI began with the publication of the policy paper Establishing a Pro-Innovation Approach to Regulating AI in 2022,[6] before the focus shifted towards frontier AI in early 2023. Most notably, the Frontier AI Taskforce was set up, which turned into the UK AI Safety Institute,[7] and the first global summit on frontier AI safety took place in Bletchley Park.[8] In the EU, the AI Act, which was finally passed in May 2024,[9] already contains specific requirements for providers of general-purpose AI models with systemic risk (e.g. to perform model evaluations, to assess and mitigate systemic risks, and to report serious incidents to the EU AI Office).[10] Against this background, the question is not whether frontier AI will be regulated, but how.

[1] DSIT, 'Frontier AI Safety Commitments, AI Seoul Summit 2024' (2024) <https://www.gov.uk/government/publications/frontier-ai-safety-commitments-ai-seoul-summit-2024> accessed 1 July 2024. This definition is a minor variation of the one set forth by DSIT, 'AI Safety Summit: Introduction' (2023) <https://www.gov.uk/government/publications/ai-safety-summit-introduction> accessed 1 July 2024. For alternative definitions, see T Shevlane and others, 'Model Evaluation for Extreme Risks' (arXiv, 2023) <http://arxiv.org/abs/2305.15324>; M Anderljung and others, 'Frontier AI Regulation: Managing Emerging Risks to Public Safety' (arXiv, 2023) <http://arxiv.org/abs/2307.03718>; MM Maas, 'Concepts in Advanced AI Governance: A Literature Review of Key Terms and Definitions' (SSRN, 2023) <https://doi.org/10.2139/ssrn.4612473>; E Jones, 'Explainer: What Is a Foundation Model?' (*Ada Lovelace Institute*, 17 July 2023) <https://www.adalovelaceinstitute.org/resource/foundation-models-explainer> accessed 1 July 2024. The definition has also been discussed in a series of workshops, H Toner and T Fist, 'Regulating the AI Frontier: Design Choices and Constraints' (*Center for Security and Emerging Technology*, 26 October 2023) <https://cset.georgetown.edu/article/regulating-the-ai-frontier-design-choices-and-constraints> accessed 1 July 2024. But note that the term has also been criticized, see G Helfrich, 'The Harms of Terminology: Why We Should Reject So-Called "Frontier AI"' (2024) AI and Ethics <https://doi.org/10.1007/s43681-024-00438-1>.

[2] The White House, 'Fact Sheet: Biden-Harris Administration Secures Voluntary Commitments from Leading Artificial Intelligence Companies to Manage the Risks Posed by AI' (2023) <https://www.whitehouse.gov/briefing-room/statements-releases/2023/07/21/fact-sheet-biden-harris-administration-secures-voluntary-commitments-from-leading-artificial-intelligence-companies-to-manage-the-risks-posed-by-ai> accessed 1 July 2024.

[3] US Senate Committee on the Judiciary, 'Oversight of AI: Rules for Artificial Intelligence' (2023) <https://www.judiciary.senate.gov/committee-activity/hearings/oversight-of-ai-rules-for-artificial-intelligence> accessed 1 July 2024; US Senate Committee on the Judiciary, 'Oversight of AI: Principles for Regulation' (2023) <https://www.judiciary.senate.gov/committee-activity/hearings/oversight-of-ai-principles-for-regulation> accessed 1 July 2024; US Senate Committee on the Judiciary, 'Oversight of AI: Legislating on Artificial Intelligence' (2023) <https://www.judiciary.senate.gov/committee-activity/hearings/oversight-of-ai-legislating-on-artificial-intelligence> accessed 1 July 2024.

This chapter focuses on the challenge of choosing a regulatory approach for frontier AI. Among other things, policymakers need to decide whether to formulate high-level principles (e.g. 'AI systems should be safe and secure') or specific rules (e.g. 'models must be fine-tuned using reinforcement learning from human feedback').[11] The corresponding regulatory approaches are typically referred to as 'principle-based regulation'[12] and 'rule-based regulation'.[13] The two approaches have complementary strengths and weaknesses. While specific rules provide more certainty and are easier to enforce, they can quickly become outdated and lead to a box-ticking attitude. Conversely, high-level principles provide less certainty and are more costly to enforce. At the same time, they are more adaptable to changes in the environment and are often more appropriate in situations where it is unclear what behavior would best advance a given regulatory objective. However, rule-based and principle-based regulation are not binaries; they exist on a spectrum. Instead of simply picking one of the two approaches, policymakers need to choose a point on the spectrum. Importantly, the appropriate level of specificity can differ significantly between individual requirements and may also change over time. Choosing the right point on the spectrum involves difficult trade-offs,[14] but it is essential for creating a regulatory regime that mitigates risks while also promoting innovation.[15]

---

[4] The White House, 'Readout of White House Meeting with CEOs on Advancing Responsible Artificial Intelligence Innovation' (2023) <https://www.whitehouse.gov/briefing-room/statements-releases/2023/05/04/readout-of-white-house-meeting-with-ceos-on-advancing-responsible-artificial-intelligence-innovation> accessed 1 July 2024.

[5] The White House, 'Safe, Secure, and Trustworthy Development and Use of Artificial Intelligence' (2023) Executive Order 14110 <https://www.federalregister.gov/documents/2023/11/01/2023-24283/safe-secure-and-trustworthy-development-and-use-of-artificial-intelligence> accessed 1 July 2024.

[6] DCMS, 'Establishing a Pro-Innovation Approach to Regulating AI' <https://www.gov.uk/government/publications/establishing-a-pro-innovation-approach-to-regulating-ai> accessed 1 July 2024.

[7] DSIT, 'Introducing the AI Safety Institute' (2023) <https://www.gov.uk/government/publications/ai-safety-institute-overview> accessed 1 July 2024. Note that other countries have set up similar institutes, see e.g. NIST, 'US AI Safety Institute' (2023) <https://www.nist.gov/aisi> accessed 1 July 2024; Japan AI Safety Institute, 'AISI: Japan AI Safety Institute' (2024) <https://aisi.go.jp> accessed 1 July 2024. The UK AI Safety Institute collaborates with some of the other institutes, see e.g. DSIT, 'UK & US Announce Partnership on Science of AI Safety' (2024) <https://www.gov.uk/government/news/uk-united-states-announce-partnership-on-science-of-ai-safety> accessed 1 July 2024.

[8] DSIT, 'AI Safety Summit: Introduction' (2023) <https://www.gov.uk/government/publications/ai-safety-summit-introduction> accessed 1 July 2024.

[9] European Parliament, 'Regulation of the European Parliament and of the Council laying down Harmonised Rules on Artificial Intelligence (Artificial Intelligence Act)' (2024) <https://data.consilium.europa.eu/doc/document/PE-24-2024-INIT/en/pdf> accessed 1 July 2024.

[10] See Articles 52a–52e of the EU AI Act.

The chapter draws from two bodies of literature. First, it is based on recent scholarship on frontier AI regulation. The most relevant source is a multi-authored report that sets out key regulatory challenges, along with building blocks for a potential regulatory regime.[16] This work has been complemented by recent discussion of specific elements of frontier AI regulation, such as licensing requirements,[17] requirements for compute providers,[18] and ways to future-proof frontier AI regulation.[19] There have also been a number of notable workshops and roundtables on the topic.[20] Second, the chapter builds upon decades of academic work on regulatory approaches,[21] which include rule-based regulation,[22] principle-based regulation,[23] risk-based regulation,[24] and meta-regulation.[25] There is also some work on regulatory approaches in an AI context. For example, there are several articles that analyze the risk-based approach in the EU AI Act,[26] while others examine the appropriateness of principle-based regulation for medical products that use AI.[27]

---

[11] For more information on the related discussion in the context of Constitutional AI, see S Kundu and others, 'Specific Versus General Principles for Constitutional AI' (arXiv, 2023) <https://arxiv.org/abs/2310.13798>.

[12] For more information, see Section II.B.1.

[13] For more information, see Section II.B.2.

[14] For more information on how to balance rule-based and principle-based approaches, see S Arjoon, 'Striking a Balance Between Rules and Principles-based Approaches for Effective Governance: A Risks-based Approach' (2006) 68 Journal of Business Ethics 53 <https://doi.org/10.1007/s10551-006-9040-6>; C Decker, 'Goals-Based and Rules-Based Approaches' (2018) BEIS Research Paper Number 8 <https://www.gov.uk/government/publications/regulation-goals-based-and-rules-based-approaches> accessed 1 July 2024; C Diver, 'The Optimal Precision of Administrative Rules' (1985) 93 Yale Law Journal 65.

[15] See B Prainsack and N Forgó, 'New AI Regulation in the EU Seeks to Reduce Risk Without Assessing Public Benefit' (2024) Nature Medicine <https://doi.org/10.1038/s41591-024-02874-2>.

[16] M Anderljung and others, 'Frontier AI Regulation: Managing Emerging Risks to Public Safety' (arXiv, 2023) <http://arxiv.org/abs/2307.03718>. For summaries of the report, see M Anderljung and A Korinek, 'Frontier AI Regulation: Safeguards Amid Rapid Progress' (*Lawfare*, 4 January 2024) <https://www.lawfaremedia.org/article/frontier-ai-regulation-safeguards-amid-rapid-progress>; M Anderljung, J Schuett and R Trager, 'Frontier AI Regulation' (*Centre for the Governance of AI*, 10 July 2023) <https://www.governance.ai/post/frontier-ai-regulation> accessed 1 July 2024.

---

[17] N Guha and others, 'AI Regulation Has Its Own Alignment Problem: The Technical and Institutional Feasibility of Disclosure, Registration, Licensing, and Auditing' (forthcoming) George Washington Law Review <https://ssrn.com/abstract=4634443>; G Smith, 'Licensing Frontier AI Development: Legal Considerations and Best Practices' (*Lawfare*, 3 January 2024) <https://www.lawfaremedia.org/article/licensing-frontier-ai-development-legal-considerations-and-best-practices> accessed 1 July 2024; D Carpenter, 'Approval Regulation for Frontier Artificial Intelligence: Pitfalls, Plausibility, Optionality' (*Harvard Kennedy School*, 2024) <https://www.hks.harvard.edu/centers/mrcbg/programs/growthpolicy/approval-regulation-frontier-artificial-intelligence-pitfalls> accessed 1 July 2024.

[18] J Egan and L Heim, 'Oversight for Frontier AI Through a Know-Your-Customer Scheme for Compute Providers' (arXiv, 2023) <https://arxiv.org/abs/2310.13625>; L Heim and others, 'Governing Through the Cloud: The Intermediary Role of Compute Providers in AI Regulation' (arXiv, 2024) <https://arxiv.org/abs/2403.08501>; G Sastry, 'Computing Power and the Governance of Artificial Intelligence' (arXiv, 2024) <https://arxiv.org/abs/2402.08797>; L Heim, M Anderljung and H Belfield, 'To Govern AI, We Must Govern Compute' (*Lawfare*, 28 March 2024) <https://www.lawfaremedia.org/article/to-govern-ai-we-must-govern-compute> accessed 1 July 2024.

[19] P Scharre, 'Future-Proofing Frontier AI Regulation: Projecting Future Compute for Frontier AI Models' (*Center for a New American Security*, 13 March 2024) <https://www.cnas.org/publications/reports/future-proofing-frontier-ai-regulation> accessed 1 July 2024; K Pilz, L Heim and N Brown, 'Increased Compute Efficiency and the Diffusion of AI Capabilities' (arXiv, 2023) <https://arxiv.org/abs/2311.15377>.

[20] H Toner and others, 'Skating to Where the Puck Is Going: Anticipating and Managing Risks From Frontier AI Systems' (*Center for Security and Emerging Technology*, October 2023) <https://cset.georgetown.edu/publication/skating-to-where-the-puck-is-going> accessed 1 July 2024; H Toner and T Fist, 'Regulating the AI Frontier: Design Choices and Constraints' (*Center for Security and Emerging Technology*, 26 October 2023) <https://cset.georgetown.edu/article/regulating-the-ai-frontier-design-choices-and-constraints> accessed 1 July 2024.

[21] For an overview of different regulatory approaches, see R Baldwin, M Cave and M Lodge, *Understanding Regulation: Theory, Strategy, and Practice* (2nd edn, Oxford University Press 2011) <https://doi.org/10.1093/acprof:osobl/9780199576081.001.0001>; R Baldwin, M Cave and M Lodge (eds), *The Oxford Handbook of Regulation* (Oxford University Press 2010) <https://doi.org/10.1093/oxfordhb/9780199560219.001.0001>; C Decker, 'Goals-Based and Rules-Based Approaches' (2018) BEIS Research Paper Number 8 <https://www.gov.uk/government/publications/regulation-goals-based-and-rules-based-approaches> accessed 1 July 2024.

---

[22] L Kaplow, 'Rules Versus Standards: An Economic Analysis' (1992) 42 Duke Law Journal 557; CR Sunstein, 'Problems With Rules' (1995) 83 California Law Review 953; C Coglianese, 'Rule Design: Defining the Regulator-Regulatee Relationship' in JC Le Coze and B Journé (eds), *The Regulator-Regulatee Relationship in High-Hazard Industry Sectors* (Springer 2024) <https://doi.org/10.1007/978-3-031-49570-0_10>.

[23] J Black, 'Forms and Paradoxes of Principles-Based Regulation' (2008) 3 Capital Markets Law Journal 425 <https://doi.org/10.1093/cmlj/kmn026>; J Black, M Hopper and C Band, 'Making a Success of Principles-Based Regulation' (2007) 1 Law and Financial Markets Review 191 <https://doi.org/10.1080/17521440.2007.11427879>; LA Cunningham, 'A Prescription to Retire the Rhetoric of "Principles-Based Systems" in Corporate Law, Securities Regulation, and Accounting' (2007) 60 Vanderbilt Law Review 1409; CL Ford, 'New Governance, Compliance, and Principles-Based Securities Regulation' (2008) 45 American Business Law Journal 1 <https://doi.org/10.1111/j.1744-1714.2008.00050.x>.

[24] J Black, 'The Emergence of Risk-Based Regulation and the New Public Management in the United Kingdom' (2005) Public Law 512; BM Hutter, 'The Attractions of Risk-Based Regulation: Accounting for the Emergence of Risk Ideas in Regulation' (*Centre for Analysis of Risk and Regulation*, 2005) <https://www.lse.ac.uk/accounting/assets/CARR/documents/D-P/Disspaper33.pdf> accessed 1 July 2024; H Rothstein and others, 'The Risks of Risk-Based Regulation: Insights From the Environmental Policy Domain' (2006) 32 Environment International 1056 <https://doi.org/10.1016/j.envint.2006.06.008>; J Black and R Baldwin, 'Really Responsive Risk-Based Regulation' (2010) 32 Law & Policy 181 <https://doi.org/10.1111/j.1467-9930.2010.00318.x>; J Black, 'Risk-Based Regulation: Choices, Practices and Lessons Being Learnt' in OECD (ed), *Risk and Regulatory Policy: Improving the Governance of Risk* (2010) <https://doi.org/10.1787/9789264082939-en>; J Black and R Baldwin, 'When Risk-Based Regulation Aims Low: Approaches and Challenges' (2012) 6 Regulation & Governance 2 <https://doi.org/10.1111/j.1748-5991.2011.01124.x>; R Baldwin and J Black, 'Driving Priorities in Risk-based Regulation: What's the Problem?' (2016) 43 Journal of Law and Society 565 <https://doi.org/10.1111/jols.12003>.

[25] R Baldwin, M Cave and M Lodge, *Understanding Regulation: Theory, Strategy, and Practice* (2nd edn, Oxford University Press 2011) <https://doi.org/10.1093/acprof:osobl/9780199576081.001.0001>; C Coglianese and E Mendelson, 'Meta-Regulation and Self-Regulation' in R Baldwin, M Cave and M Lodge (eds), *The Oxford Handbook of Regulation* (Oxford University Press 2010) <https://doi.org/10.1093/oxfordhb/9780199560219.003.0008>; S Gilad, 'It Runs in the Family: Meta-Regulation and Its Siblings' (2010) 4 Regulation & Governance 485 <https://doi.org/10.1111/j.1748-5991.2010.01090.x>; C Parker, *The Open Corporation Effective Self-regulation and Democracy* (Cambridge University Press 2002); C Parker, 'Meta-Regulation: Legal Accountability for Corporate Social Responsibility' in D Kinley (ed), *Human Rights and Corporations* (Routledge 2009) <https://doi.org/10.4324/9781315252964>; FC Simon, *Meta-Regulation in Practice: Beyond Normative Views of Morality and Rationality* (Routledge 2017).

These bodies of literature leave two important gaps. The first gap is theoretical: there is little academic discussion of the merits and limitations of different regulatory approaches in the context of frontier AI regulation. The second gap is practical: there is limited guidance on how policymakers should choose a regulatory approach. Against this background, the chapter tries to answer the question:

> How should policymakers choose a regulatory approach for frontier AI? More concretely, how specific should different requirements be at the level of legislation, regulation, and voluntary standards?

The scope of the analysis is limited in three ways. First, the chapter does not cover all types of AI regulation. Instead, it focuses on frontier AI regulation, i.e. the regulation of frontier AI systems as defined above. Note that this is a moving target. At any given time, only a handful of systems and developers can be at the frontier.[28] Second, rather than engaging with the substance of frontier AI regulation, the chapter focuses on the underlying regulatory approach. However, it does not cover all questions related to choosing a regulatory approach (e.g. whether or not the regulatory regime should be technology-neutral). Instead, it focuses on the question of how specific different requirements should be. As such, it is mainly concerned with rule-based and principle-based regulation.[29] Third, although the analysis is not tied to any specific jurisdiction, regulatory efforts in the US, UK, and EU are used as running examples.

[26] T Mahler, 'Between Risk Management and Proportionality: The Risk-Based Approach in the EU's Artificial Intelligence Act Proposal' (2022) 247 Nordic Yearbook of Law and Informatics <https://doi.org/10.53292/208f5901.38a67238>; J Chamberlain, 'The Risk-Based Approach of the European Union's Proposed Artificial Intelligence Regulation: Some Comments from a Tort Law Perspective' (2022) 14 European Journal of Risk Regulation 1 <https://doi.org/10.1017/err.2022.38>; J Schuett, 'Risk Management in the Artificial Intelligence Act' (2023) European Journal of Risk Regulation <https://doi.org/10.1017/err.2023.1>; HL Fraser and J-M Bello y Villarino, 'Acceptable Risks in Europe's Proposed AI Act: Reasonableness and Other Principles for Deciding How Much Risk Management Is Enough' (2023) European Journal of Risk Regulation <https://doi.org/10.1017/err.2023.57>; R Paul, 'European Artificial Intelligence "Trusted Throughout the World": Risk-Based Regulation and the Fashioning of a Competitive Common AI Market' (2023) Regulation & Governance <https://doi.org/10.1111/rego.12563>; M Ebers, 'Truly Risk-Based Regulation of Artificial Intelligence - How to Implement the EU's AI Act' (SSRN, 2024) <https://doi.org/10.2139/ssrn.4870387>.

[27] WG Johnson, 'Flexible Regulation for Dynamic Products? The Case of Applying Principles-Based Regulation to Medical Products Using Artificial Intelligence' (2022) 14 Law, Innovation and Technology 205 <https://doi.org/10.1080/17579961.2022.2113665>.

[28] At present, this includes GPT-4 (OpenAI, 'GPT-4 Technical Report' [arXiv, 2023] <https://arxiv.org/abs/2303.08774>), Claude 3 (Anthropic, 'The Claude 3 Model Family: Opus, Sonnet, Haiku' [2024]; <https://www-cdn.anthropic.com/de8ba9b01c9ab7cbabf5c33b80b7bbc618857627/Model_Card_Claude_3.pdf> accessed 1 July 2024), and Gemini Ultra (Google DeepMind, 'Gemini: A Family of Highly Capable Multimodal Models' [arXiv, 2023] <https://arxiv.org/abs/2312.11805>). For an overview of compute-intensive AI models, see R Rahman, D Owen and J You, 'Tracking Compute-Intensive AI Models' (*Epoch*, 5 April 2024) <https://epochai.org/blog/tracking-compute-intensive-ai-models> accessed 1 July 2024.

The chapter proceeds as follows. Section II provides background information on frontier AI regulation and regulatory approaches. Section III proposes a framework that policymakers can use to choose a regulatory approach. Section IV applies the framework to the context of frontier AI regulation. Section V concludes with a summary of the main contributions and questions for further research.

[29] We do not discuss risk-based regulation—which sometimes refers to imposing requirements in proportion to the risks a certain activity poses to society—in detail. Since frontier AI regulation aims to focus on those systems that pose the highest risks, our discussion could be framed as a way to implement a risk-based approach to AI regulation. For more information on the exclusion of risk-based regulation, see Section II.B. For a general discussion of risk-based regulation in the context of AI, see also J Black and AD Murray, 'Regulating AI and Machine Learning: Setting the Regulatory Agenda' (2019) 10 European Journal of Law and Technology; T Mahler, 'Between Risk Management and Proportionality: The Risk-Based Approach in the EU's Artificial Intelligence Act Proposal' (2022) 247 Nordic Yearbook of Law and Informatics <https://doi.org/10.53292/208f5901.38a67238>; J Chamberlain, 'The Risk-Based Approach of the European Union's Proposed Artificial Intelligence Regulation: Some Comments from a Tort Law Perspective' (2022) 14 European Journal of Risk Regulation 1 <https://doi.org/10.1017/err.2022.38>; J Schuett, 'Risk Management in the Artificial Intelligence Act' (2023) European Journal of Risk Regulation <https://doi.org/10.1017/err.2023.1>; HL Fraser and J-M Bello y Villarino, 'Acceptable Risks in Europe's Proposed AI Act: Reasonableness and Other Principles for Deciding How Much Risk Management Is Enough' (2023) European Journal of Risk Regulation <https://doi.org/10.1017/err.2023.57>; R Paul, 'European Artificial Intelligence "Trusted Throughout the World": Risk-Based Regulation and the Fashioning of a Competitive Common AI Market' (2023) Regulation & Governance <https://doi.org/10.1111/rego.12563>; ME Kaminski, 'The Developing Law of AI: A Turn to Risk Regulation' (*Lawfare*, 21 April 2023) <https://www.lawfaremedia.org/article/the-developing-law-of-ai-regulation-a-turn-to-risk-regulation> accessed 1 July 2024; M Ebers, 'Truly Risk-Based Regulation of Artificial Intelligence - How to Implement the EU's AI Act' (SSRN, 2024) <https://doi.org/10.2139/ssrn.4870387>.

## II. Setting the stage

In this section, we outline the motivation for frontier AI regulation (Section II.A) and give an overview of different regulatory approaches (Section II.B). This sets the stage for the remainder of the chapter.

### *A. The motivation for frontier AI regulation*

The motivation for regulating frontier AI systems is based on two main assumptions: that frontier AI systems pose significant risks (Section II.A.1) and that industry self-regulation will not be sufficient to reduce these risks to an acceptable level (Section II.A.2). Below, we clarify these two assumptions and sketch the contours of a potential regulatory regime that focuses specifically on frontier AI systems (Section II.A.3).

#### *1. Risks from frontier AI systems*

Frontier AI systems already pose significant risks. For example, language models can produce discriminatory or offensive outputs,[30] while image-generation models can be used to create non-consensual deepfake pornography or child abuse material.[31] Authoritarian governments have also started to use frontier AI systems to generate content that supports false narratives with the goal of undermining democratic elections,[32] while cybercriminals can use AI systems to identify vulnerabilities in software systems[33] and write personalized phishing messages.[34]

---

[30] T Bolukbasi and others, 'Man Is to Computer Programmer as Woman Is to Homemaker? Debiasing Word Embeddings' (Conference on Neural Information Processing Systems, 2016) <https://arxiv.org/abs/1607.06520>; J Zou and L Schiebinger, 'AI Can Be Sexist and Racist: It's Time to Make It Fair' (2018) 559 Nature 324 <https://doi.org/10.1038/d41586-018-05707-8>; S Cave and K Dihal, 'The Whiteness of AI' (2020) 33 Philosophy & Technology 685 <https://doi.org/10.1007/s13347-020-00415-6>; A Field and others, 'A Survey of Race, Racism, and Anti-Racism in NLP' (arXiv, 2021) <http://arxiv.org/abs/2106.11410>; N Mehrabi and others, 'A Survey on Bias and Fairness in Machine Learning' (2021) 54 ACM Computing Surveys 1 <https://doi.org/10.1145/3457607>; EM Bender and others, 'On the Dangers of Stochastic Parrots: Can Language Models Be Too Big?' (ACM Conference on Fairness, Accountability, and Transparency, 2021) <https://doi.org/10.1145/3442188.3445922>; R Bommasani and others, 'On the Opportunities and Risks of Foundation Models' (arXiv, 2021) <https://arxiv.org/abs/2108.07258>; L Weidinger and others, 'Ethical and Social Risks of Harm from Language Models' (arXiv, 2021) <https://arxiv.org/abs/2112.04359>; L Weidinger and others, 'Taxonomy of Risks Posed by Language Models' (ACM Conference on Fairness, Accountability, and Transparency, 2022) <https://doi.org/10.1145/3531146.3533088>.

[31] D Harris, 'Deepfakes: False Pornography Is Here and the Law Cannot Protect You' (2019) 17 Duke Law & Technology Review 99; M Westerlund, 'The Emergence of Deepfake Technology: A Review' (2019) 9 Technology Innovation Management Review 40 <http://doi.org/10.22215/timreview/1282>; OpenAI, 'OpenAI's Commitment to Child Safety: Adopting Safety by Design Principles' (2024) <https://openai.com/index/child-safety-adopting-sbd-principles> accessed 1 July 2024.

In the future, frontier AI systems will likely pose even greater risks. One concern is that they could provide instructions for the acquisition of biological weapons, potentially democratizing the ability to cause large-scale harm.[35]

---

[32] B Wilder and Y Vorobeychik, 'Defending Elections Against Malicious Spread of Misinformation' (AAAI Conference on Artificial Intelligence, Honolulu, 2019) <https://doi.org/10.1609/aaai.v33i01.33012213>; C Marsden, T Meyer and I Brown, 'Platform Values and Democratic Elections: How Can the Law Regulate Digital Disinformation?' (2020) 36 Computer Law & Security Review 105373 <https://doi.org/10.1016/j.clsr.2019.105373>; B Buchanan and others, 'Truth, Lies, and Automation' (*Center for Security and Emerging Technology*, 2021) <https://doi.org/10.51593/2021CA003>; E Horvitz, 'On the Horizon: Interactive and Compositional Deepfakes' (arXiv, 2022) <https://arxiv.org/abs/2209.01714>; PS Park and others, 'AI Deception: A Survey of Examples, Risks, and Potential Solutions' (arXiv, 2023) <http://arxiv.org/abs/2308.14752>; R Chowdhury, 'AI-Fuelled Election Campaigns Are Here: Where Are the Rules?' (2024) 628 Nature 237 <https://doi.org/10.1038/d41586-024-00995-9>; U Ecker and others, 'Misinformation Poses a Bigger Threat to Democracy Than You Might Think' (2024) 630 Nature 29 <https://doi.org/10.1038/d41586-024-01587-3>; K Hackenburg and H Margetts, 'Evaluating the Persuasive Influence of Political Microtargeting With Large Language Models' (2024) 121 PNAS <https://doi.org/10.1073/pnas.2403116121>.

[33] N Kaloudi and J Li, 'The AI-Based Cyber Threat Landscape: A Survey' (2020) 53 ACM Computing Surveys 1 <https://doi.org/10.1145/3372823>; B Guembe and others, 'The Emerging Threat of AI-Driven Cyber Attacks: A Review' (2022) 36 Applied Artificial Intelligence 2037254 <https://doi.org/10.1080/08839514.2022.2037254>; AJ Lohn and KA Jackson, 'Will AI Make Cyber Swords or Shields? (*Center for Security and Emerging Technology*, 2022) <https://doi.org/10.51593/2022CA002>.

[34] J Hazell, 'Spear Phishing With Large Language Models' (arXiv, 2023) <http://arxiv.org/abs/2305.06972>; M Brundage and others, 'The Malicious Use of Artificial Intelligence: Forecasting, Prevention, and Mitigation' (arXiv, 2018) <http://arxiv.org/abs/1802.07228>. But note that some scholars remain skeptical, see S Kapoor and A Narayanan, 'How to Prepare for the Deluge of Generative AI on Social Media' (*Knight First Amendment Institute*, 16 June 2023) <https://knightcolumbia.org/content/how-to-prepare-for-the-deluge-of-generative-ai-on-social-media> accessed 1 July 2024.

[35] JB Sandbrink, 'Artificial Intelligence and Biological Misuse: Differentiating Risks of Language Models and Biological Design Tools' (arXiv, 2023) <http://arxiv.org/abs/2306.13952>; EH Soice and others, 'Can Large Language Models Democratize Access to Dual-Use Biotechnology?' (arXiv, 2023) <http://arxiv.org/abs/2306.03809>; DA Boiko, R MacKnight and G Gomes, 'Emergent Autonomous Scientific Research Capabilities of Large Language Models' (arXiv, 2023) <http://arxiv.org/abs/2304.05332>; A Gopal and others, 'Will Releasing the Weights of Future Large Language Models Grant Widespread Access to Pandemic Agents?' (arXiv, 2023) <https://arxiv.org/abs/2310.18233>; RF Service, 'Could Chatbots Help Devise the Next Pandemic Virus?' (2023) 380 Science 1211 <https://doi.org/10.1126/science.adj2463>; T Patwardhan and others, 'Building an Early

Although recent risk assessments conclude that existing systems do not meaningfully increase biorisk,[36] we expect future systems to be more capable and misuse risk to be higher. Another concern is that systems will become able to deceive, persuade, and manipulate people, including their operators, making it extremely challenging to evaluate their safety.[37] A third concern is that autonomous agents might become able to create copies of themselves, acquire resources, and seek power,[38] eventually evading human oversight and control.[39] Although these and other risks are more speculative and involve

---

Warning System for LLM-Aided Biological Threat Creation' (*OpenAI*, 2024) <https://openai.com/research/building-an-early-warning-system-for-llm-aided-biological-threat-creation> accessed 1 July 2024.

[36] CA Mouton, C Lucas and E Guest, 'The Operational Risks of AI in Large-Scale Biological Attacks: A Red-Team Approach' (*RAND*, 2023) <https://doi.org/10.7249/RRA2977-1>; S Kapoor and others, 'On the Societal Impact of Open Foundation Models' (arXiv, 2024) <https://arxiv.org/abs/2403.07918>.

[37] PS Park and others, 'AI Deception: A Survey of Examples, Risks, and Potential Solutions' (arXiv, 2023) <http://arxiv.org/abs/2308.14752>; T Hagendorff, 'Deception Abilities Emerged in Large Language Models' (arXiv, 2023) <http://arxiv.org/abs/2307.16513>; META Fundamental AI Research Diplomacy Team (FAIR) and others, 'Human-Level Play in the Game of Diplomacy by Combining Language Models with Strategic Reasoning' (2022) 378 Science 1067 <https://doi.org/10.1126/science.ade9097>; J Carlsmith, 'Scheming AIs: Will AIs Fake Alignment during Training in Order to Get Power?' (arXiv, 2023) <http://arxiv.org/abs/2311.08379>; Z Kenton and others, 'Alignment of Language Agents' (arXiv, 2021) <http://arxiv.org/abs/2103.14659>; E Hubinger and others, 'Sleeper Agents: Training Deceptive LLMs that Persist Through Safety Training' (arXiv, 2024) <https://arxiv.org/abs/2401.05566>; Anthropic, 'Measuring the Persuasiveness of Language Models' (2024) <https://www.anthropic.com/news/measuring-model-persuasiveness> accessed 1 July 2024; S El-Sayed and others, 'A Mechanism-Based Approach to Mitigating Harms from Persuasive Generative AI' (arXiv, 2024) <https://arxiv.org/abs/2404.15058>; T van der Weij and others, 'AI Sandbagging: Language Models can Strategically Underperform on Evaluations' (arXiv, 2024) <https://arxiv.org/abs/2406.07358>.

[38] R Ngo, L Chan and S Mindermann, 'The Alignment Problem from a Deep Learning Perspective' (arXiv, 2023) <http://arxiv.org/abs/2209.00626>; AM Turner and P Tadepalli, 'Parametrically Retargetable Decision-Makers Tend To Seek Power' (arXiv, 2022) <http://arxiv.org/abs/2206.13477>; J Carlsmith, 'Is Power-Seeking AI an Existential Risk?' (arXiv, 2022) <http://arxiv.org/abs/2206.13353>; AM Turner and others, 'Optimal Policies Tend to Seek Power' (arXiv, 2023) <http://arxiv.org/abs/1912.01683>.

[39] M Kinniment and others, 'Evaluating Language-Model Agents on Realistic Autonomous Tasks' (arXiv, 2023) <https://arxiv.org/abs/2312.11671>; A Chan and others, 'Harms from Increasingly Agentic Algorithmic Systems' (ACM Conference on Fairness, Accountability, and Transparency, 2023) <https://doi.org/10.1145/3593013.3594033>; A Chan and others, 'Visibility Into AI Agents' (ACM Conference on Fairness, Accountability, and Transparency, 2024) <https://doi.org/10.1145/3630106.3658948>; Y Shavit and others, 'Practices for Governing Agentic AI Systems' (*OpenAI*, 2023) <https://openai.com/index/practices-for-governing-agentic-ai-systems> accessed 1 July 2024; MK Cohen and others, 'Regulating Advanced Artificial Agents' (2024) 384 Science

substantial uncertainties, they are increasingly taken seriously by companies,[40] scholars,[41] and policymakers.[42]

*2. The limits of self-regulation*

Frontier AI developers already take some measures to mitigate the above-mentioned risks. For example, they evaluate their models for dangerous capabilities,[43] fine-tune them using reinforcement learning from human

---

36 <https://doi.org/10.1126/science.adl0625>; I Gabriel and others, 'The Ethics of Advanced AI Assistants' (arXiv, 2024) <https://arxiv.org/abs/2404.16244>; N Kolt, 'Governing AI Agents' (SSRN, 2024) <https://doi.org/10.2139/ssrn.4772956>.

[40] See e.g. The White House, 'Fact Sheet: Biden-Harris Administration Secures Voluntary Commitments from Leading Artificial Intelligence Companies to Manage the Risks Posed by AI' (2023) <https://www.whitehouse.gov/briefing-room/statements-releases/2023/07/21/fact-sheet-biden-harris-administration-secures-voluntary-commitments-from-leading-artificial-intelligence-companies-to-manage-the-risks-posed-by-ai> accessed 1 July 2024; DSIT, 'Frontier AI Safety Commitments, AI Seoul Summit 2024' (2024) <https://www.gov.uk/government/publications/frontier-ai-safety-commitments-ai-seoul-summit-2024/frontier-ai-safety-commitments-ai-seoul-summit-2024> accessed 1 July 2024.

[41] See e.g. Center for AI Safety, 'Statement on AI Risk' (2023) <https://www.safe.ai/work/statement-on-ai-risk> accessed 1 July 2024; D Hendrycks, M Mazeika and T Woodside, 'An Overview of Catastrophic AI Risks' (arXiv, 2023) <http://arxiv.org/abs/2306.12001>; Y Bengio and others, 'Managing Extreme AI Risks Amid Rapid Progress' (2024) 384 Science 842 <https://doi.org/10.1126/science.adn0117>. But note that some scholars think the risks are overblown, see e.g. HS Sætra and J Danaher, 'Resolving the Battle of Short- vs. Long-Term AI Risks' (2023) AI and Ethics <https://doi.org/10.1007/s43681-023-00336-y>; Nature Editorial Board, 'Stop Talking about Tomorrow's AI Doomsday When AI Poses Risks Today' (2023) 618 Nature 885 <https://doi.org/10.1038/d41586-023-02094-7>.

[42] See e.g. DSIT, 'The Bletchley Declaration by Countries Attending the AI Safety Summit' (2023) <https://www.gov.uk/government/publications/ai-safety-summit-2023-the-bletchley-declaration/the-bletchley-declaration-by-countries-attending-the-ai-safety-summit-1-2-november-2023> accessed 1 July 2024; DSIT, 'Frontier AI: Capabilities and Risks' (2023) <https://www.gov.uk/government/publications/frontier-ai-capabilities-and-risks-discussion-paper> accessed 1 July 2024; DSIT, 'International Scientific Report on the Safety of Advanced AI' (2024) <https://www.gov.uk/government/publications/international-scientific-report-on-the-safety-of-advanced-ai> accessed 1 July 2024; DSIT, 'Seoul Ministerial Statement for Advancing AI Safety, Innovation and Inclusivity' (2024) <https://www.gov.uk/government/publications/seoul-ministerial-statement-for-advancing-ai-safety-innovation-and-inclusivity-ai-seoul-summit-2024> accessed 1 July 2024.

[43] T Shevlane and others, 'Model Evaluation for Extreme Risks' (arXiv, 2023) <http://arxiv.org/abs/2305.15324>; M Kinniment and others, 'Evaluating Language-Model Agents on Realistic Autonomous Tasks' (arXiv, 2023) <https://arxiv.org/abs/2312.11671>; M Phuong and others, 'Evaluating Frontier Models for Dangerous Capabilities' (arXiv, 2024) <https://arxiv.org/abs/2403.13793>; R Laine and others, 'Me, Myself, and AI: The

feedback (RLHF),[44] and implement safety policies for extreme risks.[45] Some developers have also made voluntary safety commitments[46] and started to coordinate on safety practices via organizations like the Frontier Model Forum (FMF),[47] the Partnership on AI (PAI),[48] and the AI Alliance.[49] Such efforts can be conceptualized as self-regulation.[50]

---

Situational Awareness Dataset (SAD) for LLMs' (arXiv, 2024) <https://arxiv.org/abs/2407.04694>.

[44] P Christiano and others, 'Deep Reinforcement Learning From Human Preferences' (arXiv, 2017) <http://arxiv.org/abs/1706.03741>; DM Ziegler and others, 'Fine-Tuning Language Models From Human Preferences' (arXiv, 2019) <http://arxiv.org/abs/1909.08593>; N Lambert and others, 'Illustrating Reinforcement Learning from Human Feedback (RLHF)' (*Hugging Face*, 9 December 2022) <https://huggingface.co/blog/rlhf> accessed 1 July 2024.

[45] Anthropic, 'Responsible Scaling Policy' (2023) <https://www.anthropic.com/news/anthropics-responsible-scaling-policy> accessed 1 July 2024; OpenAI, 'Preparedness' (2023) <https://openai.com/safety/preparedness> accessed 1 July 2024; Google DeepMind, 'Introducing the Frontier Safety Framework' (2024) <https://deepmind.google/discover/blog/introducing-the-frontier-safety-framework> accessed 1 July 2024.

[46] The White House, 'Fact Sheet: Biden-Harris Administration Secures Voluntary Commitments from Leading Artificial Intelligence Companies to Manage the Risks Posed by AI' (2023) <https://www.whitehouse.gov/briefing-room/statements-releases/2023/07/21/fact-sheet-biden-harris-administration-secures-voluntary-commitments-from-leading-artificial-intelligence-companies-to-manage-the-risks-posed-by-ai> accessed 1 July 2024; DSIT, 'Frontier AI Safety Commitments, AI Seoul Summit 2024' (2024) <https://www.gov.uk/government/publications/frontier-ai-safety-commitments-ai-seoul-summit-2024/frontier-ai-safety-commitments-ai-seoul-summit-2024> accessed 1 July 2024.

[47] Frontier Model Forum, <https://www.frontiermodelforum.org> accessed 1 July 2024.

[48] Partnership on AI, <https://partnershiponai.org> accessed 1 July 2024.

[49] AI Alliance, <https://thealliance.ai> accessed 1 July 2024.

[50] For more information on self-regulation in an AI context, see T Ferretti, 'An Institutionalist Approach to AI Ethics: Justifying the Priority of Government Regulation over Self-Regulation' (2020) 9 Moral Philosophy and Politics 239 <https://doi.org/10.1515/mopp-2020-0056>. For more information on self-regulation in general, see I Maitland, 'The Limits of Self-Regulation' (1985) 27 California Management Review 135 (1985); J Kooiman and M van Vliet, 'Self-Governance As a Mode of Societal Governance' (2000) 2 Public Management: An International Journal of Research and Theory 359 <https://doi.org/10.1080/14719030000000022>; J Black, 'Decentring Regulation: Understanding the Role of Regulation and Self-Regulation in a 'Post-Regulatory' World Get access Arrow' (2001) 54 Current Legal Problems 103 <https://doi.org/10.1093/clp/54.1.103>; LAJ Senden, 'Soft Law, Self-Regulation and Co-Regulation in European Law: Where Do They Meet?' (2005) 9 Electronic Journal of Comparative Law; C Coglianese and E Mendelson, 'Meta-Regulation and Self-Regulation' in R Baldwin, M Cave and M Lodge (eds), *The Oxford Handbook of Regulation* (Oxford University Press 2010) <https://doi.org/10.1093/oxfordhb/9780199560219.003.0008>; R Baldwin, M Cave and M Lodge, *Understanding Regulation: Theory, Strategy, and Practice*

However, there are a number of reasons why policymakers should not rely on industry self-regulation to reduce the risks from frontier AI systems. One reason is that the incentives of frontier AI developers are not fully aligned with the public interest.[51] Absent regulation, market incentives will likely push developers to underinvest in safety.[52] Another reason is that frontier AI systems produce negative externalities, i.e. they impose costs on parties other than the developers.[53] Some of these costs might be very high, potentially exceeding the developers' capacity to compensate harmed parties. Government intervention is necessary to reduce societal costs to an acceptable level or to ensure that developers internalize them. A third reason is that, due to the transformative impact that frontier AI systems will likely have on society,[54] democratic governments ought to have a say in how these systems are developed and deployed. This would be even more important if, at some point, developers succeed at building artificial general intelligence (AGI),[55] i.e. AI

---

(2nd edn, Oxford University Press 2011) <https://doi.org/10.1093/acprof:osobl/9780199576081.003.0008>.

[51] P Cihon and others, 'Corporate Governance of Artificial Intelligence in the Public Interest' (2021) 12 Information 275 <https://doi.org/10.3390/info12070275>; H Toner and T McCauley, 'AI Firms Mustn't Govern Themselves, Say Ex-Members of OpenAI's Board' (*The Economist*, 26 May 2024) <https://www.economist.com/by-invitation/2024/05/26/ai-firms-mustnt-govern-themselves-say-ex-members-of-openais-board> accessed 1 July 2024.

[52] For example, if companies think that there is a 'winner-takes-all' mechanism, they might be willing to sacrifice safety in order to 'win the race', see S Armstrong, N Bostrom and C Shulman, 'Racing to the Precipice: A Model of Artificial Intelligence Development' (2016) 31 AI & Society 201 <https://doi.org/10.1007/s00146-015-0590-y>; A Askell, M Brundage and G Hadfield, 'The Role of Cooperation in Responsible AI Development' (arXiv, 2019) <https://arxiv.org/abs/1907.04534>; W Naudé and N Dimitri, 'The Race for an Artificial General Intelligence: Implications for Public Policy' (2020) 35 AI & Society 367 <https://doi.org/10.1007/s00146-019-00887-x>.

[53] T Hagendorff, 'Blind Spots in AI Ethics' (2022) 2 AI and Ethics 851 <https://doi.org/10.1007/s43681-021-00122-8>.

[54] See e.g. B Garfinkel, 'The Impact of Artificial Intelligence' in JB Bullock and others (eds) The Oxford Handbook of AI Governance (Oxford University Press 2022) <https://doi.org/10.1093/oxfordhb/9780197579329.013.5>.

[55] A number of frontier AI developers describe their goal as building AGI, see e.g. OpenAI, 'Charter' (2020) <https://openai.com/charter> accessed 1 July 2024; S Altman, 'Planning for AGI and Beyond' (*OpenAI*, 24 February 2023) <https://openai.com/index/planning-for-agi-and-beyond> accessed 1 July 2024; Google DeepMind, 'Build AI Responsibly to Benefit Humanity' (2024) <https://deepmind.google/about> accessed 1 July 2024; M Kruppa, 'Google DeepMind CEO Says Some Form of AGI Possible in a Few Years' (*The Wall Street Journal*, 2 May 2023) <http://wsj.com/articles/google-deepmind-ceo-says-some-form-of-agi-possible-in-a-few-years-2705f452> accessed 1 July 2024; Anthropic, 'Core Views on AI Safety: When, Why, What, and How' (*Anthropic*, 8 March 2023) <https://www.anthropic.com/index/core-views-on-ai-safety> accessed 1 July 2024; A Heath, 'Mark Zuckerberg's New Goal Is Creating Artificial General Intelligence' (*The Verge*, 18 January 2024)

systems that achieve or exceed human performance across a wide range of cognitive tasks.[56] While some scholars are skeptical that AGI will ever be built,[57] surveys show that many scholars think it could happen within the coming years or decades.[58] Against this background, it is not surprising that policymakers in the US, UK, and EU are interested in regulating frontier AI systems.

### *3. The contours of frontier AI regulation*

Below, we outline three main components of a potential regulatory regime that focuses specifically on frontier AI systems. We take a brief look at the regulatory target, key requirements, and different ways to enforce them.

*Regulatory target*. Policymakers need to decide *what* should be regulated (material scope), *who* should be regulated (personal scope), *where* the regulation should apply (territorial scope), and *when* it should apply (temporal scope).[59] Defining the material scope of frontier AI regulations is particularly challenging.[60] One challenge is to ensure that the regulation applies to all AI

---

<https://www.theverge.com/2024/1/18/24042354/mark-zuckerberg-meta-agi-reorg-interview> accessed 1 July 2024.

[56] B Goertzel and C Pennachin, *Artificial General Intelligence* (Springer 2007) <https://doi.org/10.1007/978-3-540-68677-4>; B Goertzel, 'Artificial General Intelligence: Concept, State of the Art, and Future Prospects' (2014) 5 Journal of Artificial General Intelligence 1 <https://doi.org/10.2478/jagi-2014-0001>; S Altman, 'Planning for AGI and Beyond' (*OpenAI*, 24 February 2023) <https://openai.com/index/planning-for-agi-and-beyond> accessed 1 July 2024; MR Morris and others, 'Levels of AGI: Operationalizing Progress on the Path to AGI' (arXiv, 2023) <https://arxiv.org/abs/2311.02462>.

[57] M Mitchell, 'Why AI is Harder Than We Think' (arXiv, 2021) <https://arxiv.org/abs/2104.12871>; R Fjelland, 'Why General Artificial Intelligence Will Not Be Realized' (2020) 7 Humanities and Social Sciences Communications <https://doi.org/10.1057/s41599-020-0494-4>.

[58] SD Baum, B Goertzel and TG Goertzel, 'How Long Until Human-Level AI? Results from an Expert Assessment' (2011) 78 Technological Forecasting and Social Change 185 <https://doi.org/10.1016/j.techfore.2010.09.006>; VC Müller and N Bostrom, 'Future Progress in Artificial Intelligence: A Survey of Expert Opinion' in VC Müller (ed), Fundamental Issues of Artificial Intelligence (Springer 2016) <https://doi.org/10.1007/978-3-319-26485-1_33>; K Grace and others, 'When Will AI Exceed Human Performance? Evidence from AI Experts' (2018) 62 Journal of Artificial Intelligence Research 729 <https://doi.org/10.1613/jair.1.11222>; K Grace and others, 'Expert Survey on Progress in AI' (AI Impacts, August 2022) <https://aiimpacts.org/2022-expert-survey-on-progress-in-ai> accessed 1 July 2024; K Grace and others, 'Thousands of AI Authors on the Future of AI' (arXiv, 2024) <https://arxiv.org/abs/2401.02843>.

[59] See J Schuett, 'Defining the Scope of AI Regulations' (2023) 15 Law, Innovation and Technology 60 <https://doi.org/10.1080/17579961.2023.2184135>.

[60] See M Anderljung and others, 'Frontier AI Regulation: Managing Emerging Risks to Public Safety' (arXiv, 2023) <http://arxiv.org/abs/2307.03718>; H Toner and T Fist, 'Regulating the AI Frontier: Design Choices and Constraints' (*Center for Security and*

systems that pose significant risks (otherwise the regulation would be under-inclusive), without also applying to systems that do not pose significant risks (otherwise the regulation would be over-inclusive).[61] But since it is often difficult to draw this line, more cautious regulatory regimes may want to include more systems (to avoid false negatives), while more permissive regimes may include fewer systems (to avoid false positives). Another challenge is to ensure that regulatees are able to determine whether or not the regulation applies to their systems *before* they need to comply with certain requirements, which may be before, during, or after training a model. For example, since model capabilities can emerge unintentionally and unpredictably during training,[62] it makes little sense to impose pre-training requirements for systems that have certain capabilities. A third challenge is to define the regulatory target in a future-proof way.[63] Since it is often not

---

*Emerging Technology*, 26 October 2023) <https://cset.georgetown.edu/article/regulating-the-ai-frontier-design-choices-and-constraints> accessed 1 July 2024; P Scharre, 'Future-Proofing Frontier AI Regulation: Projecting Future Compute for Frontier AI Models' (*Center for a New American Security*, 13 March 2024) <https://www.cnas.org/publications/reports/future-proofing-frontier-ai-regulation> accessed 1 July 2024.

[61] For more information on the over-inclusiveness and under-inclusiveness of AI regulation, see MU Scherer, 'Regulating Artificial Intelligence Systems: Risks, Challenges, Competencies, and Strategies' (2016) 29 Harvard Journal of Law & Technology 353 <https://doi.org/10.2139/ssrn.2609777>; C Reed, 'How Should We Regulate Artificial Intelligence?' (2018) 376 Philosophical Transactions of the Royal Society A 1 <https://doi.org/10.1098/rsta.2017.0360>; R Martinez, 'Artificial Intelligence: Distinguishing between Types & Definitions' (2019) 19 Nevada Law Journal 1015; B Casey and MA Lemley, 'You Might Be a Robot' (2019) 105 Cornell Law Review 287; MC Buiten, 'Towards Intelligent Regulation of Artificial Intelligence' (2019) 10 European Journal of Risk Regulation 41 <https://doi.org/10.1017/err.2019.8>; R Crootof and BJ Ard, 'Structuring Techlaw' (2021) 34 Harvard Journal of Law & Technology 366 <https://doi.org/10.2139/ssrn.3664124>; J Schuett, 'Defining the Scope of AI Regulations' (2023) 15 Law, Innovation and Technology 60 <https://doi.org/10.1080/17579961.2023.2184135>.

[62] D Ganguli and others, 'Predictability and Surprise in Large Generative Models' (ACM Conference on Fairness, Accountability, and Transparency, Seoul, 2022) <https://doi.org/10.1145/3531146.3533229>; J Wei and others, 'Emergent Abilities of Large Language Models' (arXiv, 2022) <http://arxiv.org/abs/2206.07682>. But note that some scholars remain skeptical, see R Schaeffer, B Miranda and S Koyejo, 'Are Emergent Abilities of Large Language Models a Mirage?' (Conference on Neural Information Processing Systems, 2023) <http://arxiv.org/abs/2304.15004>.

[63] P Scharre, 'Future-Proofing Frontier AI Regulation: Projecting Future Compute for Frontier AI Models' (*Center for a New American Security*, 13 March 2024) <https://www.cnas.org/publications/reports/future-proofing-frontier-ai-regulation> accessed 1 July 2024; M Anderljung and A Korinek, 'Frontier AI Regulation: Safeguards Amid Rapid Progress' (*Lawfare*, 4 January 2024) <https://www.lawfaremedia.org/article/frontier-ai-regulation-safeguards-amid-rapid-

possible to anticipate technological changes, the target may need to be adjusted over time.[64] At the moment, the most popular approach to define the material scope of frontier AI regulations is to use compute thresholds, i.e. thresholds based on the amount of computational resources used to train a model.[65] For example, the EU AI Act uses a threshold of $10^{25}$ floating-point operations (FLOP) to identify general-purpose AI models with systemic risk,[66] while the Executive Order on Safe, Secure, and Trustworthy AI uses a slightly higher threshold of $10^{26}$ operations.[67] However, it is worth noting that compute is only a proxy for model capabilities which, in turn, are only a proxy for risk. Compute thresholds are therefore only used as an initial filter to narrow down the regulatory scope. Developers of models that exceed the compute threshold are then required to conduct extensive risk assessments. The outcome of these assessments determines whether developers are subject to additional requirements.[68]

*Key requirements*. It is still unclear what requirements should be imposed on frontier AI developers. The following practices have been suggested in the literature[69] and are supported by some policymakers.[70] Among other things,

---

progress> accessed 1 July 2024; K Pilz, L Heim and N Brown, 'Increased Compute Efficiency and the Diffusion of AI Capabilities' (arXiv, 2023) <https://arxiv.org/abs/2311.15377>.

[64] C Winter and C Bullock, 'The Governance Misspecification Problem' (forthcoming).

[65] M Pistillo and others, 'The Role of Compute Thresholds for AI Governance' (forthcoming); L Heim and L Koessler, 'Training Compute Thresholds: Features and Functions in AI Regulation' (forthcoming).

[66] Articles 51(2), 55(1) of the EU AI Act.

[67] The White House, 'Safe, Secure, and Trustworthy Development and Use of Artificial Intelligence' (2023) Executive Order 14110 <https://www.federalregister.gov/documents/2023/11/01/2023-24283/safe-secure-and-trustworthy-development-and-use-of-artificial-intelligence> accessed 1 July 2024.

[68] L Koessler and others, 'Risk Thresholds for Frontier AI' (arXiv, 2024) <https://arxiv.org/abs/2406.14713>; M Pistillo and others, 'The Role of Compute Thresholds for AI Governance' (forthcoming); L Heim and L Koessler, 'Training Compute Thresholds: Features and Functions in AI Regulation' (forthcoming).

[69] For an overview of different requirements, see M Anderljung and others, 'Frontier AI Regulation: Managing Emerging Risks to Public Safety' (arXiv, 2023) <http://arxiv.org/abs/2307.03718>; H Toner and T Fist, 'Regulating the AI Frontier: Design Choices and Constraints' (*Center for Security and Emerging Technology*, 26 October 2023) <https://cset.georgetown.edu/article/regulating-the-ai-frontier-design-choices-and-constraints> accessed 1 July 2024.

[70] See e.g. DSIT, 'Emerging Processes for Frontier AI Safety' (2023) <https://www.gov.uk/government/publications/emerging-processes-for-frontier-ai-safety> accessed 1 July 2024; European Parliament, 'Regulation of the European Parliament and of the Council laying down Harmonised Rules on Artificial Intelligence (Artificial Intelligence Act)' (2024) <https://data.consilium.europa.eu/doc/document/PE-24-2024-INIT/en/pdf> accessed 1 July 2024; The White House, 'Safe, Secure, and Trustworthy Development and Use of Artificial Intelligence' (2023) Executive Order 14110

frontier AI developers could be required to: (1) conduct risk assessments (e.g. evaluate models for dangerous capabilities[71] or estimate the impact and likelihood of key risks[72]), (2) implement certain safety and security measures (e.g. cyber defenses,[73] watermarking,[74] or misuse monitoring[75]), (3) ensure that

---

<https://www.federalregister.gov/documents/2023/11/01/2023-24283/safe-secure-and-trustworthy-development-and-use-of-artificial-intelligence> accessed 1 July 2024; NIST, 'Artificial Intelligence Risk Management Framework (AI RMF 1.0)' (2023) <https://doi.org/10.6028/NIST.AI.100-1>; NIST, 'Artificial Intelligence Risk Management Framework: Generative Artificial Intelligence Profile (NIST AI 600-1) – Initial Public Draft' (2024) <https://airc.nist.gov/docs/NIST.AI.600-1.GenAI-Profile.ipd.pdf> accessed 1 July 2024.

[71] T Shevlane and others, 'Model Evaluation for Extreme Risks' (arXiv, 2023) <http://arxiv.org/abs/2305.15324>; M Kinniment and others, 'Evaluating Language-Model Agents on Realistic Autonomous Tasks' (arXiv, 2023) <https://arxiv.org/abs/2312.11671>; M Phuong and others, 'Evaluating Frontier Models for Dangerous Capabilities' (arXiv, 2024) <https://arxiv.org/abs/2403.13793>; R Laine and others, 'Me, Myself, and AI: The Situational Awareness Dataset (SAD) for LLMs' (arXiv, 2024) <https://arxiv.org/abs/2407.04694>.

[72] J Schuett and others, 'How to Estimate the Impact and Likelihood of Risks From AI' (forthcoming).

[73] S Nevo and others, 'Securing Artificial Intelligence Model Weights' (*RAND*, 2024) <https://doi.org/10.7249/RRA2849-1>; Anthropic, 'Frontier Model Security' (2023) <https://www.anthropic.com/news/frontier-model-security> accessed 1 July 2024; OpenAI, 'Reimagining Secure Infrastructure for Advanced AI' (2024) <https://openai.com/index/reimagining-secure-infrastructure-for-advanced-ai> accessed 1 July 2024.

[74] J Kirchenbauer and others, 'A Watermark for Large Language Models' (International Conference on Machine Learning, 2023) <https://arxiv.org/abs/2301.10226>; X Zhao and others, 'Protecting Language Generation Models via Invisible Watermarking' (International Conference on Machine Learning, 2023) <https://arxiv.org/abs/2302.03162>; M Saberi and others, 'Robustness of AI-Image Detectors: Fundamental Limits and Practical Attacks' (arXiv, 2023) <https://arxiv.org/abs/2310.00076>; Google DeepMind, 'Watermarking AI-Generated Text and Video With SynthID' (2024) <https://deepmind.google/discover/blog/watermarking-ai-generated-text-and-video-with-synthid> accessed 1 July 2024.

[75] M Brundage and others, 'The Malicious Use of Artificial Intelligence: Forecasting, Prevention, and Mitigation' (arXiv, 2018) <https://arxiv.org/abs/1802.07228>; M Anderljung and J Hazell, 'Protecting Society From AI Misuse: When Are Restrictions on Capabilities Warranted?' (arXiv, 2023) <https://arxiv.org/abs/2303.09377>; B Clifford, 'Preventing AI Misuse: Current Techniques' (*Centre for the Governance of AI*, 17 December 2023) <https://www.governance.ai/post/preventing-ai-misuse-current-techniques> accessed 1 July 2024; RV Yampolskiy, 'On Monitorability of AI' (2024) AI and Ethics <https://doi.org/10.1007/s43681-024-00420-x>; N Marchal and others, 'Generative AI Misuse: A Taxonomy of Tactics and Insights from Real-World Data' (arXiv, 2024) <http://arxiv.org/abs/2406.13843>.

the level of risk is below certain thresholds (e.g. capability or risk thresholds[76]), (4) subject their systems to external scrutiny (e.g. by third-party auditors[77] or independent red teams[78]), (5) disclose certain information about their systems (e.g. via system cards,[79] model cards,[80] or data sheets[81]), about their safety measures (e.g. via safety cases[82]), or about safety and security incidents (e.g. via incident databases[83] or special information-sharing regimes with government[84]), and (6) establish sound risk governance (e.g. a central risk function, a board-level risk-committee, and internal assurance[85]). Some of these requirements are discussed in more detail below.[86] We also refer to the literature on voluntary safety and governance practices[87] and relevant company policies.[88]

---

[76] L Koessler and others, 'Risk Thresholds for Frontier AI' (arXiv, 2024) <https://arxiv.org/abs/2406.14713>.

[77] ID Raji and J Buolamwini, 'Actionable Auditing: Investigating the Impact of Publicly Naming Biased Performance Results of Commercial AI Products' (AAAI/ACM Conference on AI, Ethics, and Society, 2019) <https://doi.org/10.1145/3306618.3314244>; G Falco and others, 'Governing AI Safety Through Independent Audits' (2021) 3 Nature Machine Intelligence 566 <https://doi.org/10.1038/s42256-021-00370-7>; ID Raji and others, 'Outsider Oversight: Designing a Third Party Audit Ecosystem for AI Governance' (AAAI/ACM Conference on AI, Ethics, and Society, 2022) <https://doi.org/10.1145/3514094.3534181>; J Mökander and L Floridi, 'Operationalising AI Governance Through Ethics-Based Auditing: An Industry Case Study' (2022) AI and Ethics <https://doi.org/10.1007/s43681-022-00171-7>; J Mökander and others, 'Auditing Large Language Models: A Three-Layered Approach' (2023) AI and Ethics <https://doi.org/10.1007/s43681-023-00289-2>; M Anderljung and others, 'Towards Publicly Accountable Frontier LLMs: Building an External Scrutiny Ecosystem under the ASPIRE Framework' (arXiv, 2023) <https://arxiv.org/abs/2311.14711>; A Birhane and others, 'AI auditing: The Broken Bus on the Road to AI Accountability' (arXiv, 2024) <https://arxiv.org/abs/2401.14462>.

[78] D Ganguli and others, 'Red Teaming Language Models to Reduce Harms: Methods, Scaling Behaviors, and Lessons Learned' (arXiv, 2022) <https://arxiv.org/abs/2209.07858>; E Perez and others, 'Red Teaming Language Models with Language Models' (arXiv, 2022) <https://arxiv.org/abs/2202.03286>; B Radharapu and others, 'AART: AI-Assisted Red-Teaming with Diverse Data Generation for New LLM-Powered Applications' (arXiv, 2023) <https://arxiv.org/abs/2311.08592>; CA Mouton, C Lucas and E Guest, 'The Operational Risks of AI in Large-Scale Biological Attacks: A Red-Team Approach' (*RAND*, 2023) <https://doi.org/10.7249/RRA2977-1>; Anthropic, 'Frontier Threats Red Teaming for AI Safety' (2023) <https://www.anthropic.com/news/frontier-threats-red-teaming-for-ai-safety> accessed 1 July 2024; Anthropic, 'Challenges in Red Teaming AI Systems' (2024) <https://www.anthropic.com/news/challenges-in-red-teaming-ai-systems> accessed 1 July 2024; M Samvelyan, 'Rainbow Teaming: Open-Ended Generation of Diverse Adversarial Prompts' (arXiv, 2024) <https://arxiv.org/abs/2402.16822>; L Weidinger and others, 'STAR: SocioTechnical Approach to Red Teaming Language Models' (arXiv, 2024) <https://arxiv.org/abs/2406.11757>.

[79] N Green and others, 'System Cards: A New Resource for Understanding How AI Systems Work' (*Meta AI*, 23 February 2022) <https://ai.meta.com/blog/system-cards-a-new-resource-for-understanding-how-ai-systems-work> accessed 1 July 2024.

[80] M Mitchell and others, 'Model Cards for Model Reporting' (Conference on Fairness, Accountability, and Transparency, 2019) <https://doi.org/10.1145/3287560.3287596>.

[81] T Gebru and others, 'Datasheets for Datasets' (2021) 64 Communications of the ACM 86 <https://doi.org/10.1145/3458723>; E Bender and B Friedman, 'Data Statements for Natural Language Processing: Toward Mitigating System Bias and Enabling Better Science' (2018) 6 Transactions of the Association for Computational Linguistics 587 <https://doi.org/10.1162/tacl_a_00041>.

*Enforcement*. There are different ways in which these requirements could be enforced. A useful distinction can be drawn between enforcement measures that sanction past non-compliant behavior (*ex-post*) and measures aimed at future non-compliant behavior (*ex-ante*). *Ex-post* measures might include imposing tort liability[89] or administrative fines on developers or taking systems

---

[82] J Clymer and others, 'Safety Cases: How to Justify the Safety of Advanced AI Systems' (arXiv, 2024) <https://arxiv.org/abs/2403.10462>; A Wasil and others, 'Affirmative Safety: An Approach to Risk Management for Advanced AI' (SSRN, 2024) <https://doi.org/10.2139/ssrn.4806274>; MD Buhl and others, 'Safety Cases for Frontier AI' (forthcoming).

[83] S McGregor, 'Preventing Repeated Real World AI Failures by Cataloguing Incidents: The AI Incident Database' (AAAI Conference on Artificial Intelligence, 2021) <https://doi.org/10.1609/aaai.v35i17.17817>; OECD, 'AI Incidents Monitor' (2024) <https://oecd.ai/en/incidents> accessed 1 July 2024.

[84] N Kolt and others, 'Responsible Reporting for Frontier AI Development' (arXiv, 2024) <https://arxiv.org/abs/2404.02675>; N Mulani and J Whittlestone, 'Proposing a Foundation Model Information-Sharing Regime for the UK' (*Centre for the Governance of AI*, 16 June 2023) <https://www.governance.ai/post/proposing-a-foundation-model-information-sharing-regime-for-the-uk> accessed 1 July 2024.

[85] J Schuett, 'Three Lines of Defense Against Risks From AI' (2023) AI & Society <https://doi.org/10.1007/s00146-023-01811-0>; J Schuett, 'Frontier AI Developers Need an Internal Audit Function' (arXiv, 2023) <https://arxiv.org/abs/2305.17038>.

[86] See Section IV.A.

[87] E.g. J Schuett and others, 'Towards Best Practices in AGI Safety and Governance: A Survey of Expert Opinion' (arXiv, 2023) <https://arxiv.org/abs/2305.07153>.

[88] E.g. Anthropic,'Responsible Scaling Policy' (2023) <https://www.anthropic.com/news/anthropics-responsible-scaling-policy> accessed 1 July 2024; OpenAI, 'Preparedness' (2023) <https://openai.com/safety/preparedness> accessed 1 July 2024; Google DeepMind, 'Introducing the Frontier Safety Framework' (2024) <https://deepmind.google/discover/blog/introducing-the-frontier-safety-framework> accessed 1 July 2024.

[89] See e.g. MU Scherer, 'Regulating Artificial Intelligence Systems: Risks, Challenges, Competencies, and Strategies' (2016) 29 Harvard Journal of Law & Technology 353 <https://doi.org/10.2139/ssrn.2609777>; J Chamberlain, 'The Risk-Based Approach of the European Union's Proposed Artificial Intelligence Regulation: Some Comments from a Tort Law Perspective' (2022) 14 European Journal of Risk Regulation 1 <https://doi.org/10.1017/err.2022.38>; G Weil, 'Tort Law as a Tool for Mitigating Catastrophic Risk from Artificial Intelligence' (SSRN, 2024) <https://doi.org/10.2139/ssrn.4694006>; M van der Merwe and others, 'Tort Law and

from the market,[90] whereas *ex-ante* measures might include the need to obtain a license,[91] submit a safety case,[92] or purchase insurance.[93] A discussion of the benefits and limitations of different measures is beyond the scope of this chapter. And although we do not take a stand on which enforcement measures are most appropriate, we expect that a combination of multiple measures will be needed.

Finally, we wish to emphasize that frontier AI regulation would only be a part of a much broader regulatory ecosystem. In this chapter, we are not able to cover other important aspects, such as the international dimension of frontier

---

Frontier AI Governance' (*Lawfare*, 24 May 2024) <https://www.lawfaremedia.org/article/tort-law-and-frontier-ai-governance> accessed 1 July 2024.

[90] M Anderljung and others, 'Frontier AI Regulation: Managing Emerging Risks to Public Safety' (arXiv, 2023) <http://arxiv.org/abs/2307.03718>.

[91] M Anderljung and others, 'Frontier AI Regulation: Managing Emerging Risks to Public Safety' (arXiv, 2023) <http://arxiv.org/abs/2307.03718>; G Smith, 'Licensing Frontier AI Development: Legal Considerations and Best Practices' (*Lawfare*, 3 January 2024) <https://www.lawfaremedia.org/article/licensing-frontier-ai-development-legal-considerations-and-best-practices> accessed 1 July 2024; D Carpenter, 'Approval Regulation for Frontier Artificial Intelligence: Pitfalls, Plausibility, Optionality' (*Harvard Kennedy School*, 2024) <https://www.hks.harvard.edu/centers/mrcbg/programs/growthpolicy/approval-regulation-frontier-artificial-intelligence-pitfalls> accessed 1 July 2024.

[92] See J Clymer and others, 'Safety Cases: How to Justify the Safety of Advanced AI Systems' (arXiv, 2024) <https://arxiv.org/abs/2403.10462>; A Wasil and others, 'Affirmative Safety: An Approach to Risk Management for Advanced AI' (SSRN, 2024) <https://doi.org/10.2139/ssrn.4806274>; MD Buhl and others, 'Safety Cases for Frontier AI' (forthcoming).

[93] See MG Faure and S Li, 'Artificial Intelligence and (Compulsory) Insurance' (2022) 13 Journal of European Tort Law 1 <https://doi.org/10.1515/jetl-2022-0001>; A Gopal and others, 'Will Releasing the Weights of Future Large Language Models Grant Widespread Access to Pandemic Agents?' (arXiv, 2023) <https://arxiv.org/abs/2310.18233>. For more information on mandatory liability insurance more generally, see T Baker and R Swedloff, 'Regulation by Liability Insurance: From Auto to Lawyers Professional Liability' (2013) 60 UCLA Law Review 1412; MG Faure, 'The Complementary Roles of Liability, Regulation and Insurance in Safety Management: Theory and Practice' (2014) 17 Journal of Risk Research 689 <https://doi.org/10.1080/13669877.2014.889199>; KS Abraham and DB Schwarcz, 'The Limits of Regulation by Insurance' (2022) 98 Indiana Law Journal 215.

AI regulation[94] and the regulation of applications based on frontier AI systems,[95] among other things.

*B. Overview of different regulatory approaches*

In the following, we give an overview of two common regulatory approaches: rule-based regulation (Section II.B.1) and principle-based regulation (Section II.B.2).[96] We briefly describe both approaches (Table 1) and discuss their relative advantages and disadvantages (Table 2). Note that the two approaches are compatible with risk-based regulation, which is another regulatory approach.[97] Since frontier AI systems pose significant risks,[98] a regulatory regime that attempts to reduce these risks at least implicitly takes a risk-based approach.

[94] P Cihon and others, 'Fragmentation and the Future: Investigating Architectures for International AI Governance' (2020) 11 Global Policy 545 <https://doi.org/10.1111/1758-5899.12890>; OJ Erdélyi and J Goldsmith, 'Regulating Artificial Intelligence: Proposal for a Global Solution' (2022) 39 Government Information Quarterly <https://doi.org/10.1016/j.giq.2022.101748>; L Ho and others, 'International Institutions for Advanced AI' (arXiv, 2023) <https://arxiv.org/abs/2307.04699>; R Trager and others, 'International Governance of Civilian AI: A Jurisdictional Certification Approach' (arXiv, 2023) <https://arxiv.org/abs/2308.15514>; MM Maas and JJ Villalobos, 'International AI Institutions: A Literature Review of Models, Examples, and Proposals' (SSRN, 2023) <https://doi.org/10.2139/ssrn.4579773>; H Roberts and others, 'Global AI Governance: Barriers and Pathways Forward' (2024) 10 International Affairs 1275 <https://doi.org/10.1093/ia/iiae073>.

[95] See e.g. M Viljanen and H Parviainen, 'AI Applications and Regulation: Mapping the Regulatory Strata' (2022) 3 Frontiers in Computer Science 1 <https://doi.org/10.3389/fcomp.2021.779957>; P Hacker and others, 'Regulating ChatGPT and other Large Generative AI Models' (ACM Conference on Fairness, Accountability, and Transparency, 2023) <https://doi.org/10.1145/3593013.3594067>.

[96] For an overview of different approaches to AI regulation, see N Petit and J De Cooman. 'Models of Law and Regulation for AI' in A Elliott (ed), *The Routledge Social Science Handbook of AI* (Routledge 2021) <https://doi.org/10.4324/9780429198533>. For a taxonomy of approaches to the regulation of advanced AI that is focused more on distinct regulating actors and their available regulatory levers, see MM Maas, 'Advanced AI Governance: A Literature Review of Problems, Options, and Proposals' (SSRN, 2023) <https://doi.org/10.2139/ssrn.4629460>.

---

[97] The term 'risk-based regulation' has different meanings. It can mean imposing requirements in proportion to the risks a certain activity poses to society. But it can also mean regulation that is developed using risk management tools, such as risk assessment techniques and cost-benefit-analysis—essentially approaching regulation like a business activity. For more information on risk-based regulation in general, see J Black, 'The Emergence of Risk-Based Regulation and the New Public Management in the United Kingdom' (2005) Public Law 512; BM Hutter, 'The Attractions of Risk-Based Regulation: Accounting for the Emergence of Risk Ideas in Regulation' (*Centre for Analysis of Risk and Regulation*, 2005) <https://www.lse.ac.uk/accounting/assets/CARR/documents/D-P/Disspaper33.pdf> accessed 1 July 2024; H Rothstein and others, 'The Risks of Risk-Based Regulation: Insights From the Environmental Policy Domain' (2006) 32 Environment International 1056 <https://doi.org/10.1016/j.envint.2006.06.008>; J Black and R Baldwin, 'Really Responsive Risk-Based Regulation' (2010) 32 Law & Policy 181 <https://doi.org/10.1111/j.1467-9930.2010.00318.x>; J Black, 'Risk-Based Regulation: Choices, Practices and Lessons Being Learnt' in OECD (ed), *Risk and Regulatory Policy: Improving the Governance of Risk* (2010) <https://doi.org/10.1787/9789264082939-en>; J Black and R Baldwin, 'When Risk-Based Regulation Aims Low: Approaches and Challenges' (2012) 6 Regulation & Governance 2 <https://doi.org/10.1111/j.1748-5991.2011.01124.x>; R Baldwin and J Black, 'Driving Priorities in Risk-based Regulation: What's the Problem?' (2016) 43 Journal of Law and Society 565 <https://doi.org/10.1111/jols.12003>. For more information on risk-based regulation in the context of AI, see J Black and AD Murray, 'Regulating AI and Machine Learning: Setting the Regulatory Agenda' (2019) 10 European Journal of Law and Technology; T Mahler, 'Between Risk Management and Proportionality: The Risk-Based Approach in the EU's Artificial Intelligence Act Proposal' (2022) 247 Nordic Yearbook of Law and Informatics <https://doi.org/10.53292/208f5901.38a67238>; J Chamberlain, 'The Risk-Based Approach of the European Union's Proposed Artificial Intelligence Regulation: Some Comments from a Tort Law Perspective' (2022) 14 European Journal of Risk Regulation 1 <https://doi.org/10.1017/err.2022.38>; J Schuett, 'Risk Management in the Artificial Intelligence Act' (2023) European Journal of Risk Regulation <https://doi.org/10.1017/err.2023.1>; HL Fraser and J-M Bello y Villarino, 'Acceptable Risks in Europe's Proposed AI Act: Reasonableness and Other Principles for Deciding How Much Risk Management Is Enough' (2023) European Journal of Risk Regulation <https://doi.org/10.1017/err.2023.57>; R Paul, 'European Artificial Intelligence "Trusted Throughout the World": Risk-Based Regulation and the Fashioning of a Competitive Common AI Market' (2023) Regulation & Governance <https://doi.org/10.1111/rego.12563>; ME Kaminski, 'The Developing Law of AI: A Turn to Risk Regulation' (*Lawfare*, 21 April 2023) <https://www.lawfaremedia.org/article/the-developing-law-of-ai-regulation-a-turn-to-risk-regulation> accessed 1 July 2024; M Ebers, 'Truly Risk-Based Regulation of Artificial Intelligence - How to Implement the EU's AI Act' (SSRN, 2024) <https://doi.org/10.2139/ssrn.4870387>.

[98] See Section II.A.1.

| | Rule-based regulation | Principle-based regulation |
|---|---|---|
| How specific are the requirements? | Rules are more specific; they prescribe or prohibit a specific behavior | Principles are more abstract; they do not prescribe or prohibit a specific behavior |
| Who decides on the content of the requirements? | Policymakers draft the rules | Regulatees interpret the principles and decide how to best comply with them |
| When is the content of the requirements determined? | When policymakers draft the rules | When regulatees interpret the principles and take action |
| How is alignment between the requirements and regulatory objectives ensured? | Policymakers attempt to draft rules such that, if regulatees comply with them, the regulatory objectives will be achieved | Regulatees are expected to interpret principles and take action in a way that is aligned with the regulatory objectives |
| How is compliance with the requirements assessed? | Regulators check whether regulatees have complied with the rules | Regulators check whether regulatees' behavior is consistent with the principles |
| How much discretion do regulators have? | Regulators typically have limited discretion | Regulators typically have significant discretion |

*Table 1*: Conceptual differences between pure versions[99] of rule-based and principle-based regulation[100]

### *1. Rule-based regulation*

The basic idea behind rule-based regulation is that policymakers attempt to formulate specific rules that, if followed, would achieve a given regulatory objective.[101] Rules typically prescribe or prohibit a specific behavior. For example, a speed limit prohibits a specific driving behavior to advance the objective of safe motorways.[102] Since rules are precisely drafted and highly particularistic, regulatees can determine in advance (*ex-ante*) what behavior is permissible (e.g. not driving faster than 70 miles per hour on a motorway). Compliance is assessed by comparing the regulatees' behavior to the rules (e.g. via speed controls). This process is mostly mechanical with limited exceptions and flexibility. A potential rule in the context of frontier AI regulation could be 'before releasing a frontier AI system, it must be evaluated for dangerous model capabilities'.

[99] Note that, in practice, hybrid versions are much more common, see Section III.A.1.

[100] Based on C Decker, 'Goals-Based and Rules-Based Approaches' (2018) BEIS Research Paper Number 8 <https://www.gov.uk/government/publications/regulation-goals-based-and-rules-based-approaches> accessed 1 July 2024.

[101] For more information on rules-based regulation, see C Decker, 'Goals-Based and Rules-Based Approaches' (2018) BEIS Research Paper Number 8 <https://www.gov.uk/government/publications/regulation-goals-based-and-rules-based-approaches> accessed 1 July 2024; L Kaplow, 'Rules Versus Standards: An Economic Analysis' (1992) 42 Duke Law Journal 557; CR Sunstein, 'Problems With Rules' (1995) 83 California Law Review 953; C Coglianese, 'Rule Design: Defining the Regulator-Regulatee Relationship' in JC Le Coze and B Journé (eds), *The Regulator-Regulatee Relationship in High-Hazard Industry Sectors* (Springer 2024) <https://doi.org/10.1007/978-3-031-49570-0_10>.

Rule-based regulation is the most common type of regulation. It is used across many different industries and sectors. For example, the document Biosafety in Microbiological and Biomedical Laboratories (BMBL), published by the US Centers for Disease Control and Prevention (CDC) and the US National Institutes of Health (NIH), defines 'Biosafety Levels' for biological agents that laboratories may deal with.[103] For each Biosafety Level, the regulated laboratories are required to take specific containment measures (e.g. how to prevent unauthorized access to the laboratory, or where to place sinks for handwashing). Another example is US nuclear regulation (though it moved towards a more principle-based approach over time).[104] The US Nuclear Regulatory Commission (NRC) issues regulations and detailed guidelines that operators of nuclear power plants need to follow to obtain a license.[105]

### *2. Principle-based regulation*

The basic idea behind principle-based regulation is that, instead of prescribing a set of specific behaviors, policymakers formulate high-level principles.[106]

---

[102] C Decker, 'Goals-Based and Rules-Based Approaches' (2018) BEIS Research Paper Number 8 <https://www.gov.uk/government/publications/regulation-goals-based-and-rules-based-approaches> accessed 1 July 2024; CR Sunstein, 'Problems With Rules' (1995) 83 California Law Review 953.

[103] CDC and NIH, 'Biosafety in Microbiological and Biomedical Laboratories (BMBL) 6th Edition' (2020) <https://www.cdc.gov/labs/BMBL.html> accessed 1 July 2024. Note that the Biosafety Levels were the blueprint for Anthropic's AI Safety Levels (ASLs), see Anthropic, 'Responsible Scaling Policy' (2023) <https://www.anthropic.com/news/anthropics-responsible-scaling-policy> accessed 1 July 2024.

[104] NRC, 'History of the NRC's Risk-Informed Regulatory Programs' (2021) <https://www.nrc.gov/about-nrc/regulatory/risk-informed/history.html> accessed 1 July 2024; TR Wellock, *Safe Enough? A History of Nuclear Power and Accident Risk* (University of California Press 2021).

[105] NRC, 'Regulations, Guidance, and Communications' (2022) <https://www.nrc.gov/reactors/operator-licensing/regs-guides-comm.html> accessed 1 July 2024.

[106] For more information on principle-based regulation, see J Black, 'Forms and Paradoxes of Principles-Based Regulation' (2008) 3 Capital Markets Law Journal 425

These principles are then interpreted by the regulatees who have to decide on their own how to best comply with them. For example, instead of setting a specific speed limit, drivers could be required to 'drive at a reasonable speed'.[107] A possible principle in the context of frontier AI regulation could be that 'frontier AI systems should be safe and secure'. Compliance is assessed by evaluating whether the regulatees' behavior is consistent with the principles, which is determined on a case-by-case basis. Principle-based regulation has also been referred to as performance-based regulation,[108]

---

<https://doi.org/10.1093/cmlj/kmn026>; J Black, M Hopper and C Band, 'Making a Success of Principles-Based Regulation' (2007) 1 Law and Financial Markets Review 191 <https://doi.org/10.1080/17521440.2007.11427879>; C Decker, 'Goals-Based and Rules-Based Approaches' (2018) BEIS Research Paper Number 8 <https://www.gov.uk/government/publications/regulation-goals-based-and-rules-based-approaches> accessed 1 July 2024; PJ May, 'Performance-Based Regulation and Regulatory Regimes: The Saga of Leaky Buildings' (2003) 25 Law & Policy 381 <https://doi.org/10.1111/j.0265-8240.2003.00155.x>; PJ May, 'Regulatory Regimes and Accountability' (2007) 1 Regulation & Governance 8 <https://doi.org/10.1111/j.1748-5991.2007.00002.x>; E Seger, 'In Defence of Principlism in AI Ethics and Governance' (2022) 35 Philosophy & Technology <https://doi.org/10.1007/s13347-022-00538-y>.

[107] C Decker, 'Goals-Based and Rules-Based Approaches' (2018) BEIS Research Paper Number 8 <https://www.gov.uk/government/publications/regulation-goals-based-and-rules-based-approaches> accessed 1 July 2024; CR Sunstein, 'Problems With Rules' (1995) 83 California Law Review 953.

[108] C Coglianese, J Nash and T Olmstead, 'Performance-Based Regulation: Prospects and Limitations in Health, Safety, and Environmental Protection' (2003) 55 Administrative Law Review 705; C Coglianese, 'Performance-Based Regulation: Concepts and Challenges' in F Bignami and D Zaring (eds), *Comparative Law and Regulation: Understanding the Global Regulatory Process Research Handbooks in Comparative Law series* (Edward Elgar Publishing 2016) <https://doi.org/10.4337/9781782545613.00028>; C Coglianese, 'The Limits of Performance-Based Regulation' (2017) 50 University of Michigan Journal of Law Reform 525 <https://doi.org/10.36646/mjlr.50.3.limits>; LE Willis, 'Performance-Based Remedies: Ordering Firms to Eradicate Their Own Fraud' (2017) 80 Law and Contemporary Problems 7; PJ May, 'Performance-Based Regulation and Regulatory Regimes: The Saga of Leaky Buildings' (2003) 25 Law & Policy 381 <https://doi.org/10.1111/j.0265-8240.2003.00155.x>; SD Sugarman, 'Salt, High Blood Pressure, and Performance-Based Regulation' (2009) 3 Regulation & Governance 84 <https://doi.org/10.1111/j.1748-5991.2009.01048.x>; R Deighton-Smith, 'Process and Performance-Based Regulation: Challenges for Regulatory Governance and Regulatory Reform' in P Carroll, R Deighton-Smith, H Silver and C Walker (eds), *Minding the Gap: Appraising the Promise and Performance of Regulatory Reform in Australia* (ANU Press 2008) <https://doi.org/10.26530/OAPEN_459375>.

outcome-based regulation,[109] goal-based regulation,[110] and management-based regulation.[111] While some scholars consider these labels to be largely synonymous, others think that they differ substantively.[112] In any case, the approaches all share the feature of being less specific than rule-based regulation.

Principle-based regulation has also become very popular in many regulatory domains. A common example is crash testing for cars, where manufacturers are required to meet broad safety objectives rather than following specific testing procedures. Another example is UK financial services law, where firms must show that fair treatment of customers is at the heart of their business model.[113]

| | Rule-based regulation | Principle-based regulation |
|---|---|---|
| Need to know adequate behavior | It is difficult to draft rules without knowing what behavior would be adequate. | Principles can be drafted even if it is still unclear what behavior would be adequate. |
| Burden to identify adequate behavior | Burden is put on policymakers, who may or may not have the necessary expertise. | Burden is put on regulatees, who are more likely to have the necessary expertise. |

[109] J Smith, K Ross and H Whiley, 'Australian Food Safety Policy Changes from a "Command and Control" to an "Outcomes-Based" Approach: Reflection on the Effectiveness of Its Implementation' (2006) 13 International Journal of Environmental Research and Public Health 1218 <https://doi.org/10.3390/ijerph13121218>.

[110] C Decker, 'Goals-Based and Rules-Based Approaches' (2018) BEIS Research Paper Number 8 <https://www.gov.uk/government/publications/regulation-goals-based-and-rules-based-approaches> accessed 1 July 2024.

[111] C Coglianese and D Lazer, 'Management-Based Regulation: Prescribing Private Management to Achieve Public Goals' (2003) 37 Law & Society Review 691 <https://doi.org/10.1046/j.0023-9216.2003.03703001.x>.

[112] C Decker, 'Goals-Based and Rules-Based Approaches' (2018) BEIS Research Paper Number 8 <https://www.gov.uk/government/publications/regulation-goals-based-and-rules-based-approaches> accessed 1 July 2024.

[113] J Black, 'Forms and Paradoxes of Principles-Based Regulation' (2008) 3 Capital Markets Law Journal 425 <https://doi.org/10.1093/cmlj/kmn026>; J Black, M Hopper and C Band, 'Making a Success of Principles-Based Regulation' (2007) 1 Law and Financial Markets Review 191 <https://doi.org/10.1080/17521440.2007.11427879>; CA Petit and T Becker, 'Recent Trends in UK Financial Sector Regulation and Possible Implications for the EU, Including Its Approach to Equivalence' (*European Parliament*, 2023) <https://www.europarl.europa.eu/thinktank/en/document/IPOL_STU(2023)740067> accessed 1 July 2024.

| Over- and under-inclusiveness[114] | Rules are often over- or under-inclusive. | Principles are usually not over- or under-inclusive. |
| --- | --- | --- |
| Accountability | Policymakers are responsible for ensuring that rules are aligned with the regulatory objective. | The responsibility is partly shifted towards regulatees, which can lead to accountability gaps. |
| Need to update | Rules that govern emerging technologies need to be updated more frequently, which can be difficult.[115] | Principles are more future-proof and need to be updated less frequently. |

[114] Rules or principles are over-inclusive if they include cases which are not in need of regulation according to the regulatory objective. They are under-inclusive if cases which should have been included are not included, see R Baldwin, M Cave and M Lodge, *Understanding Regulation: Theory, Strategy, and Practice* (2nd edn, Oxford University Press 2011) <https://doi.org/10.1093/acprof:osobl/9780199576081.001.0001>; R Crootof and BJ Ard, 'Structuring Techlaw' (2021) 34 Harvard Journal of Law & Technology 347 <https://doi.org/10.2139/ssrn.3664124>.

[115] For more information on future-proofing regulations, see DD Friedman, 'Does Technology Require New Law?' (2001) 25 Harvard Journal of Law & Public Policy 71; LB Moses, 'Recurring Dilemmas: The Law's Race to Keep up with Technological Change' (2007) 2 Journal of Law, Technology & Policy 239; A Thompson, 'Rational Design in Motion: Uncertainty and Flexibility in the Global Climate Regime' (2010) 16 European Journal of International Relations 269 <https://doi.org/10.1177/1354066109342918>; GE Marchant, 'Addressing the Pacing Problem' in GE Marchant, BR Allenby and JR Herkert (eds), *The Growing Gap Between Emerging Technologies and Legal-Ethical Oversight: The Pacing Problem* (Springer 2011) <https://doi.org/10.1007/978-94-007-1356-7_13>; W Wallach, *A Dangerous Master How to Keep Technology from Slipping Beyond Our Control* (Basic Books 2015); LB Moses, 'Regulating in the Face of Sociotechnical Change' in R Brownsword, E Scotford and K Yeung (eds), *The Oxford Handbook of Law, Regulation and Technology* (Oxford University Press 2016), <https://doi.org/10.1093/oxfordhb/9780199680832.013.49>; R Brownsword, E Scotford and K Yeung, 'Law, Regulation, and Technology: The Field, Frame, and Focal Questions' in R Brownsword, E Scotford and K Yeung (eds), *The Oxford Handbook of Law, Regulation and Technology* (Oxford University Press 2016) <https://doi.org/10.1093/oxfordhb/9780199680832.013.1>; MU Scherer, 'Regulating Artificial Intelligence Systems: Risks, Challenges, Competencies, and Strategies' (2016) 29 Harvard Journal of Law & Technology 353 <https://doi.org/10.2139/ssrn.2609777>; GE Marchant and YA Stevens, 'Resilience: A New Tool in the Risk Governance Toolbox for Emerging Technologies' (2017) 51 UC Davis Law Review 233; A Chander, 'Future-Proofing Law' (2017) 51 UC Davis Law Review 1 <https://ssrn.com/abstract=3065023>; R Crootof and BJ Ard, 'Structuring Techlaw' (2021) 34 Harvard Journal of Law & Technology 347 <https://doi.org/10.2139/ssrn.3664124>; MC Buiten, 'Towards Intelligent Regulation of Artificial Intelligence' (2019) 10 European Journal of Risk Regulation 41 <https://doi.org/10.1017/err.2019.8>; C Reed, 'How Should We Regulate Artificial Intelligence?' (2018) 376 Philosophical Transactions of the Royal Society A 1 <https://doi.org/10.1098/rsta.2017.0360>; B Casey and MA Lemley, 'You Might Be a

| | | |
|---|---|---|
| Legal certainty | Rules provide more legal certainty. | Principles provide less legal certainty. They can lead to regulatory bias or arbitrariness. |
| Incentives for innovation | Rules provide limited incentives for regulatees to develop innovative compliance approaches. | Principles encourage regulatees to innovate on compliance (e.g. to develop novel safety measures). |
| Incentives for compliance | Regulatees might adopt a box-ticking attitude or even try to 'game the rules' and exploit loopholes.[116] | Principles can lead to over- or under-compliance, depending on the regulatees' risk profiles. |
| Difficulty of verifying compliance | It is typically easier to verify compliance with rules ('has the regulatee followed the rules?'). | It is typically more difficult to verify compliance with principles ('is the regulatee's behavior consistent with the principles?'). |
| Differential treatment of regulatees | Formally, all regulatees are treated the same. | Principles can allow for differential treatment of regulatees based on their compliance history or other factors. |
| Enforcement costs | Enforcing rules is often less costly, mainly because they do not have to make case-by-case decisions or because the adequate behavior is relatively straightforward and uncontested. | Enforcing principles is often more costly. |

*Table 2*: Relative advantages and disadvantages of rule-based and principle-

Robot' (2019) 105 Cornell Law Review 287; MM Maas, 'Innovation-Proof Governance for Military AI?' (2019) 10 Journal of International Humanitarian Legal Studies 129 <https://doi.org/10.1163/18781527-01001006>.

[116] The Volkswagen Diesel Emission scandal provides a stark illustration of this dynamic, see JC Jung and E Sharon, 'The Volkswagen Emissions Scandal and Its Aftermath' (2019) 38 Global Business and Organizational Excellence 6 <https://doi.org/10.1002/joe.21930>.

based regulation[117]

## III. How to choose a regulatory approach

In this section, we propose a framework that policymakers can use to choose a regulatory approach. More concretely, the framework helps to answer the question: how specific should different requirements be at the level of legislation, regulation, and voluntary standards (Figure 1)? Below, we describe the elements of the framework (Section III.A), and explain how to apply it (Section III.B).

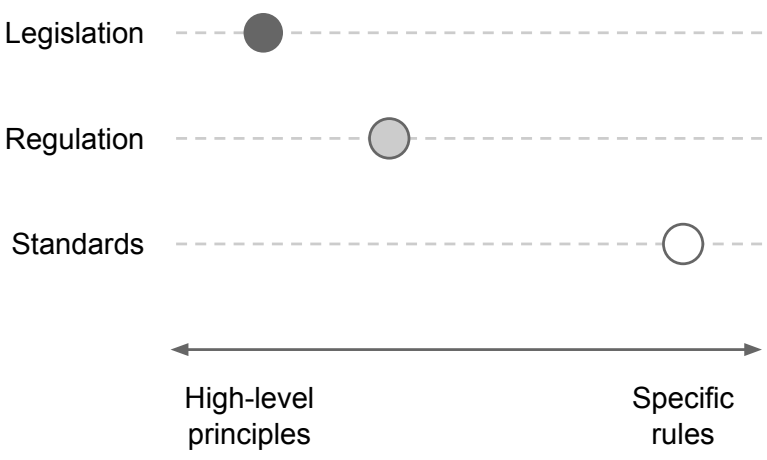

*Figure 1*: Framework for deciding how specific requirements should be at the level of legislation, regulation, and voluntary standards

[117] Based on C Decker, 'Goals-Based and Rules-Based Approaches' (2018) BEIS Research Paper Number 8 <https://www.gov.uk/government/publications/regulation-goals-based-and-rules-based-approaches> accessed 1 July 2024. For more information about the advantages and disadvantages of rule-based regulation, see L Kaplow, 'Rules Versus Standards: An Economic Analysis' (1992) 42 Duke Law Journal 557; CR Sunstein, 'Problems With Rules' (1995) 83 California Law Review 953; S Arjoon, 'Striking a Balance Between Rules and Principles-based Approaches for Effective Governance: A Risks-based Approach' (2006) 68 Journal of Business Ethics 53 <https://doi.org/10.1007/s10551-006-9040-6>; C Winter and C Bullock, 'The Governance Misspecification Problem' (forthcoming). For more information about the advantages and disadvantages of principle-based regulation, see LA Cunningham, 'A Prescription to Retire the Rhetoric of "Principles-Based Systems" in Corporate Law, Securities Regulation, and Accounting' (2007) 60 Vanderbilt Law Review 1409; R Deighton-Smith, 'Process and Performance-Based Regulation: Challenges for Regulatory Governance and Regulatory Reform' in P Carroll, R Deighton-Smith, H Silver and C Walker (eds), *Minding the Gap: Appraising the Promise and Performance of Regulatory Reform in Australia* (ANU Press 2008) <https://doi.org/10.26530/OAPEN_459375>.

### *A. Elements of the framework*

The framework has two main elements: a spectrum that measures a requirement's level of specificity (Section III.A.1) and different actors who can specify the requirement (Section III.A.2).[118]

#### *1. Level of specificity*

Principle-based and rule-based regulation are not dichotomous. They are located on a spectrum. On one end of the spectrum, requirements are formulated as high-level principles that regulatees have to interpret. On the other end of the spectrum, requirements are formulated as specific rules that prescribe or prohibit a specific behavior. Crucially, the appropriate level of specificity can differ significantly between different requirements within a broader regime. Talking about the specificity of an entire regulatory regime would therefore be an oversimplification. As such, our framework distinguishes between the specificity of individual requirements (or clusters of requirements), rather than the regulatory regime as a whole.[119]

The appropriate level of specificity of different requirements may also change over time (Figure 2). Sometimes, requirements are initially more abstract (e.g. because it is unclear what behavior best advances the regulatory objective) and gradually become more specific (e.g. because best practices emerge). Other times, requirements may become more principle-based. For example, in the US nuclear industry, regulators initially imposed specific requirements on power plants intended to keep the risk of a reactor meltdown to an acceptable level.[120] However, from the 1990s onwards, as it became clear that the requirements were not able to prevent major accidents (e.g. the Three Mile Island accident),[121] while still being considered overly burdensome, the Nuclear Regulatory Commission (NRC) moved towards a more principle-based regime.[122]

[118] Note that we ignore other relevant dimensions (e.g. how requirements are enforced or how voluntary vs. binding they are).

[119] See LA Cunningham, 'A Prescription to Retire the Rhetoric of "Principles-Based Systems" in Corporate Law, Securities Regulation, and Accounting' (2007) 60 Vanderbilt Law Review 1409.

[120] TR Wellock, *Safe Enough? A History of Nuclear Power and Accident Risk* (University of California Press 2021); NRC, 'History of the NRC's Risk-Informed Regulatory Programs' (2021) <https://www.nrc.gov/about-nrc/regulatory/risk-informed/history.html> accessed 1 July 2024.

[121] See e.g. A Hopkins, 'Was Three Mile Island a "Normal Accident"?' (2002) 9 Journal of Contingencies and Crisis Management 65 <https://doi.org/10.1111/1468-5973.00155>.

[122] Note that there is an important disanalogy between the history of nuclear regulation and frontier AI regulation: since nuclear technology was first developed by the US military, regulators initially had more relevant expertise than industry. This is certainly not the case with regards to AI.

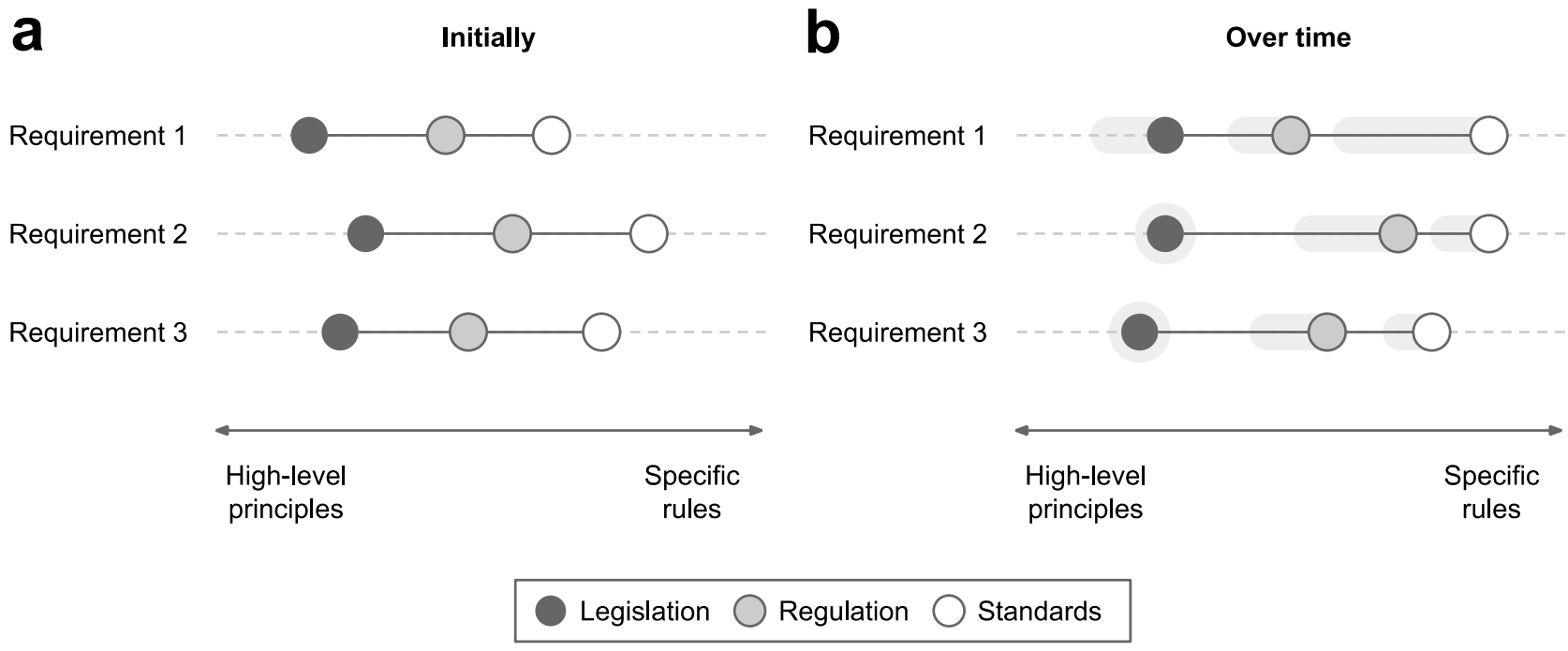

*Figure 2*: Illustrating that the appropriate degree of specificity may change over time

*2. Actors who can specify requirements*

The appropriate degree of specificity depends on the level in the hierarchy of law, i.e. the ranking of different sources of law. In most legal systems, some legal sources are higher-ranking than others, and higher-ranking sources take priority over lower-ranking sources (*lex superior derogat legi inferiori*). For example, in the US and the EU (but not in the UK), constitutional law is superior to acts of parliament. Relatedly, higher-ranking sources are typically more abstract than lower-ranking sources, especially where lower-ranking sources specify the substance of high-ranking sources (e.g. regulations may specify acts of parliament).[123] Against this background, policymakers need to decide how specific different requirements should be at different levels in the hierarchy of law.

For the purposes of our framework, we distinguish between three levels in the hierarchy of law: (1) legislation, (2) regulation, and (3) voluntary standards. This is clearly a simplification and some levels are missing (e.g. constitutional law and other administrative rules). Since we do not focus on a specific jurisdiction, it will also not reflect details that might be relevant in some jurisdictions. Below, we explain each of the three levels.

[123] There has been a parallel discussion on how to turn AI ethics principles into practice, see B Mittelstadt, 'Principles Alone Cannot Guarantee Ethical AI' (2019) 1 Nature Machine Intelligence 501 <https://doi.org/10.1038/s42256-019-0114-4>; J Morley and others, 'From What to How: An Initial Review of Publicly Available AI Ethics Tools, Methods and Research to Translate Principles Into Practices' (2020) 26 Science and Engineering Ethics 2142 <https://doi.org/10.1007/s11948-019-00165-5>; L Floridi, 'Translating Principles Into Practices of Digital Ethics: Five Risks of Being Unethical' (2021) 32 Philosophy & Technology 185 <https://doi.org/10.1007/s13347-019-00354-x>; J Zhou and F Chen, 'AI Ethics: From Principles to Practice' (2022) 38 AI & Society 2693 <https://doi.org/10.1007/s00146-022-01602-z>; E Seger, 'In Defence of Principlism in AI Ethics and Governance' (2022) 35 Philosophy & Technology 45 <https://doi.org/10.1007/s13347-022-00538-y>.

*Legislation*. By 'legislation', we mean laws adopted by a legislator through the legislative process. At the federal level in the US, these are Acts of Congress. In the UK, these are Acts of Parliament. In the EU, this consists of Directives (e.g. the proposed AI Liability Directive), which EU member states have to transpose into national law,[124] and Regulations (e.g. the EU AI Act), which are directly applicable in all EU member states.[125] In the US and UK, this type of law is often referred to as 'primary legislation', but note that the term has a different meaning in the EU.[126]

*Regulation*. By 'regulation', we mean administrative rules issued by an executive agency or regulatory body. Although US executive orders are also issued by the executive, namely the president, they serve different functions and follow different processes for issuance and implementation. For our purposes, we still qualify them as regulation. In the UK, regulations are also called 'secondary legislation'. In the EU, this includes Delegated Acts, which enable the European Commission to supplement or amend non-essential parts of EU legislation,[127] and Implementing Acts, which lay down detailed rules allowing for their uniform implementation.[128] Recall that EU Regulations are a type of legislation passed by the European Parliament and the Council; they are not regulation in our sense.

*Voluntary standards*. By 'voluntary standards', we mean technical or operational specifications developed by a standard-setting body. This body can be a private organization (e.g. the International Organization for Standardization [ISO] or the International Electrotechnical Commission [IEC]), an organization appointed by law (e.g. the European Standards Organizations [ESO] or the British Standards Institution [BSI]), or even a state actor (e.g. the US National Institute of Standards and Technology [NIST]). Unlike regulations, standards are voluntary. Examples include the NIST AI Risk Management Framework,[129] the harmonized standards on risk management that are currently being developed by CEN-CENELEC,[130] and ISO/IEC 23894.[131] Standards are typically created and amended in a multi-stakeholder standard-setting process. In the EU, standards developed by appointed standard-setting bodies have a special effect: if regulatees comply with the standard, it is presumed that they all comply with the regulation or legislation. This effect is called 'presumption of conformity'.

---

[124] See Article 288 of the Treaty on the Functioning of the European Union (TFEU).

[125] See Article 288 of the TFEU.

[126] In the EU, the term 'primary legislation' refers to the international treaties between the EU member states that provides the constitutional basis for the EU. It includes the Treaty on European Union (TEU) and the Treaty on the Functioning of the European Union (TFEU).

[127] See Article 290 of the TFEU.

[128] See Article 291 of the TFEU.

[129] NIST, 'Artificial Intelligence Risk Management Framework (AI RMF 1.0)' (2023) <https://doi.org/10.6028/NIST.AI.100-1>. See also NIST, 'Artificial Intelligence Risk Management Framework: Generative Artificial Intelligence Profile (NIST AI 600-1)' (2024) <https://airc.nist.gov/docs/NIST.AI.600-1.GenAI-Profile.ipd.pdf> accessed 1 July 2024.

Requirements at lower levels in the hierarchy of law are always more specific than requirements at higher levels. Since regulations add specificity to legislation, and voluntary standards add specificity to regulations, voluntary standards will always be more specific than the regulations they build upon, and regulations more specific than the legislation they build on. As a result, one rarely finds highly specific legislation or highly abstract standards (but there are also exceptions). In our framework, the appropriate level of specificity is therefore skewed to the right for lower levels in the hierarchy of law. In other words, the order is always black, gray, white.

This framework can be used to illustrate the difference between rule-based and principle-based regulation. Under a rule-based approach, requirements remain somewhat abstract at the legislative level, while they are fairly specific at the level of regulation and voluntary standards. Under a principle-based approach, requirements at the legislative and regulatory level remain abstract, while they might only be a bit more specific at the level of voluntary standards. Figure 3 illustrates the differences. Points represent the medium level of specificity across different requirements. The range shows the level of specificity of the majority of requirements, though there may be outliers.

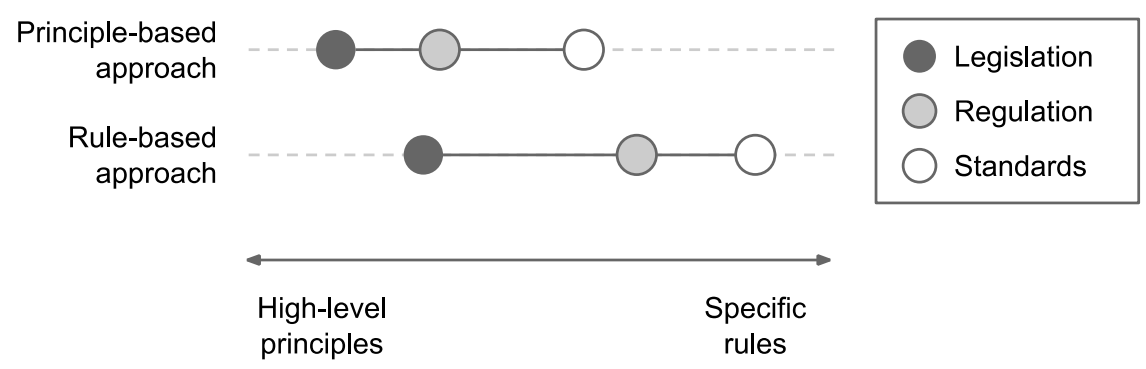

*Figure 3*: Illustrating the difference between principle-based and rule-based regulation

### *B. How to apply the framework*

To apply the framework, we propose two sets of questions. The first set of questions can be used to determine how specific a requirement should be at most (Section III.B.1). This sets the outer boundary to the right on the spectrum. Going beyond that boundary would result in overly specific rules that may impose excessive regulatory burdens or fail to meet their regulatory objective as a result of misspecification. The second set of questions can then be used to determine who should specify the requirement (Section III.B.2). It helps to allocate the responsibility for turning high-level principles into more specific rules between the legislator, the regulator, and standard-setting bodies. Figure 4 illustrates the purpose of the two sets of questions. Most of the questions are based on the literature on regulatory approaches,[132] though we have adapted them to fit our framework, and added new questions where appropriate. Table 3 summarizes the questions.

[130] J Soler Garrido and others, 'Analysis of the Preliminary AI Standardisation Work Plan in Support of the AI Act' (*European Commission*, 2023) <https://doi.org/10.2760/5847>.

[131] ISO/IEC 23894:2023 Information Technology – Artificial Intelligence – Guidance on Risk Management <https://www.iso.org/standard/77304.html> accessed 1 July 2024.

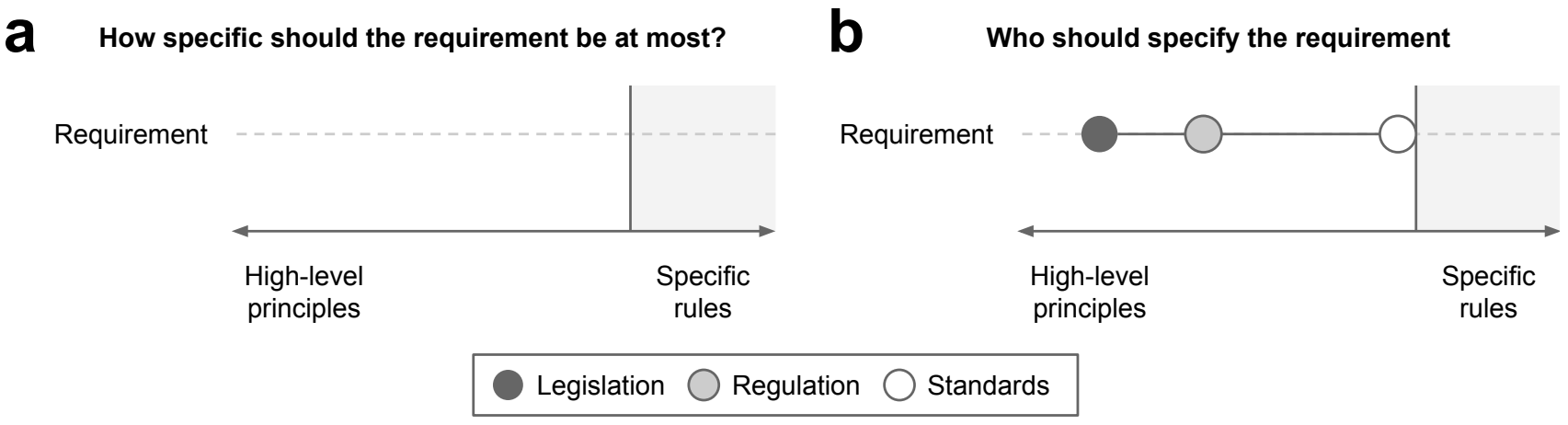

*Figure 4*: Illustrating the purpose of the two sets of questions

| How specific should the requirement be at most? | Who should specify the requirement? |
|---|---|
| • How well-understood are relevant risks?<br>• Is it clear what specific behavior would reduce those risks?<br>• How important is legal certainty?<br>• How aligned are the regulatees' incentives with the regulatory objective? | • How much expertise and information do the actors have?<br>• How often do the requirements need to be updated and how onerous is the updating process?<br>• How aligned are the actors' incentives with the regulatory objective?<br>• To what extent are higher-level actors able to oversee lower-level actors? |

*Table 3*: Overview of the two sets of questions

### *1. How specific should the requirement be at most?*

The following considerations can be used to determine the maximum degree of specificity of individual requirements:

[132] E.g. C Decker, 'Goals-Based and Rules-Based Approaches' (2018) BEIS Research Paper Number 8 <https://www.gov.uk/government/publications/regulation-goals-based-and-rules-based-approaches> accessed 1 July 2024.

*How well-understood are relevant risks?* In general, requirements should only be formulated as highly specific rules if relevant risks are well-understood. If the risks are poorly understood, specific rules intended to address those risks can easily be ineffective or become ineffective as the risk landscape evolves. Specific rules can also impose an unnecessary compliance burden on regulatees (e.g. if the risks are lower than initially expected). While it is certainly possible to reduce poorly-understood risks via specific rules, such rules will often be less effective or more onerous than they need to be. The following questions can be used to determine how well-understood relevant risks are: How much do we know about current risks? Do we have a mechanistic understanding of the chain of events that lead to harm ('risk model')? Can we predict future risks? How quickly is the risk landscape evolving? How much do experts agree and how confident are they in their views?

*Is it clear what specific behavior would reduce those risks?* Requirements should only prescribe or prohibit a specific behavior if it is clear that the behavior would actually reduce relevant risks. If this is not clear, it will often be preferable to formulate high-level principles instead. To determine how well-understood a risk-reducing behavior is, policymakers can use the following considerations: Do best practices exist? How complex is the behavior? How context-specific is it? How diverse are the different contexts? How different are the regulatees? How often do regulatees have to perform the behavior? How quickly does it change what behavior is considered responsible? Does the behavior have spillover effects?

*How important is legal certainty?* Legal certainty is generally desirable (e.g. because it can incentivize companies to make longer-term investments). However, legal certainty often trades off against other regulatory objectives (e.g. reducing risk). Policymakers therefore need to decide how important legal certainty is relative to these other objectives. As a general rule of thumb, the more important legal certainty is, the more specific the requirement should be. This is because assessing compliance with specific rules is often easier than assessing compliance with high-level principles. However, ease of compliance is not only determined by the requirements' level of specificity. For example, companies might still need to spend significant resources to assess compliance with a long list of very technical requirements, even if the requirements are formulated as specific rules. Conversely, they might find it relatively straightforward to comply with some principles. Against this background, it is not always clear whether legal certainty favors specific rules or high-level principles. Other relevant considerations include: How many actors does the regime apply to? How difficult is it to assess whether regulatees have complied with high-level principles? Are there clear tests for doing so?

*How aligned are the regulatees' incentives with the regulatory objective?* If the regulatees' incentives are already aligned with the regulatory objective, the requirements can be less specific and grant the regulatees more flexibility. This is because regulatees' will likely interpret the requirements in a way that advances the regulatory objective. Conversely, if the regulatees' interests diverge significantly from the public interest (e.g. if their ordinary practice produces significant negative externalities), it will often be preferable to prescribe or prohibit a specific behavior, or ensure considerable oversight over how regulatees implement principles. Other relevant considerations include: Are there considerable externalities? To what extent do regulatees already adhere to voluntary standards? Do regulatees appear to do the bare minimum to comply with existing requirements or do they behave more responsibly than legally required? How much are regulatees expected to 'game the rules' and exploit loopholes? Do the regulatees appear to have a considerably higher risk appetite than society at large?

*2. Who should specify the requirement?*

The following considerations can be used to allocate the responsibility for specifying the requirement between the legislator, regulators, and standard-setting bodies:

*How much expertise and information do different actors have?* In general, requirements should be specified by actors who have relevant domain expertise and access to relevant information. To avoid misspecification, actors who do not have the necessary expertise and information should generally not formulate highly specific rules. Note that it also matters how much expertise and information actors have relative to other actors, not just how much they have in absolute terms. Another important consideration is how much expertise they can build up and how much information they can obtain in the future.

*How often do the requirements need to be updated and how onerous is the updating process?* If the requirements need to be updated more frequently, they should be specified by actors who can update them more easily. Updating requirements at the legislative level usually takes the longest. Updating voluntary standards can also be rather slow, especially if the updating process involves many different stakeholders and requires consensus. Whether it takes longer to update requirements at the regulatory or sub-regulatory level varies by jurisdiction.

*How aligned are the actors' incentives with the regulatory objective?* Requirements should generally be specified by actors whose incentives are aligned with the regulatory objective. For example, if a regulator mainly cares about avoiding accidents, without taking into account costs to innovation, policymakers may prefer that more of the specification be done by standard-setting bodies. Similarly, if standard-setting bodies are dominated by industry actors whose primary goal is to reduce regulatory uncertainty and compliance costs, while neglecting negative externalities, more of the specification work should be left in the hands of the regulator.

*To what extent are higher-level actors able to oversee lower-level actors?* Higher-level actors (e.g. regulators) often delegate the task of specifying the requirements to a lower-level actor (e.g. standard-setting bodies). However, they may only want to do this if they can oversee the lower-level actor or if they can reverse the delegation. For example, regulators might be hesitant to hand over important details to a standard-setting body if the body is dominated by industry interests, the standard-setting process is opaque, and the delegation is irreversible. However, there are ways in which higher-level actors can maintain some amount of influence. For example, regulators can reserve the power to determine whether compliance with a voluntary standard presumes conformity with certain regulations.

## IV. Choosing a regulatory approach for frontier AI

In this section, we apply our framework to the context of frontier AI regulation. We use the practices listed in the policy paper Emerging Processes for Frontier AI Safety that the UK Department for Science, Innovation and Technology (DSIT) published ahead of the AI Safety Summit 2023.[133] For each of the nine practices, we use our questions to determine how specific they should be at the level of legislation, regulation, and voluntary standards (Section IV.A). We then discuss the implications of our analysis and make policy recommendations (Section IV.B).

### *A. Applying our framework to nine AI safety practices*

DSIT's list represents the most comprehensive and detailed list of safety practices to date.[134] Although the practices are voluntary, treating the list as legal requirements helps to illustrate our framework. It is worth noting, however, that the practices set forth by DSIT are fairly specific. This level of specificity implies a more rule-based approach. To avoid a bias in that direction, we try to describe the practices as abstractly as possible. But even then, more extreme versions of a principle-based approach, where different areas of safety practices are not even mentioned, are effectively ruled out. We also assume that work on the different safety practices will continue. For example, we assume that legislation will get passed and new standards will be developed. Finally, we wish to emphasize that our framework is a rather coarse tool. We have substantial uncertainties about the precise positions of many points.

[133] DSIT, 'Emerging Processes for Frontier AI Safety' (2023) <https://www.gov.uk/government/publications/emerging-processes-for-frontier-ai-safety> accessed 1 July 2024.

[134] Other lists of frontier AI safety practices have been proposed by M Anderljung and others, 'Frontier AI Regulation: Managing Emerging Risks to Public Safety' (arXiv, 2023) <http://arxiv.org/abs/2307.03718>; J Schuett and others, 'Towards Best Practices in AGI Safety and Governance: A Survey of Expert Opinion' (arXiv, 2023) <https://arxiv.org/abs/2305.07153>; Anthropic, 'Responsible Scaling Policy' (2023) <https://www.anthropic.com/news/anthropics-responsible-scaling-policy> accessed 1 July 2024; OpenAI, 'Preparedness' (2023) <https://openai.com/safety/preparedness> accessed 1 July 2024.

DSIT's policy paper lists nine practices: responsible capability scaling (Section IV.A.1), model evaluations and red-teaming (Section IV.A.2), model-reporting and information-sharing (Section IV.A.3), security controls including securing model weights (Section IV.A.4), reporting structure for vulnerabilities (Section IV.A.5), identifiers of AI-generated material (Section IV.A.6), prioritizing research on risks posed by AI (Section IV.A.7), preventing and monitoring model misuse (Section IV.A.8), and data input controls and audits (Section IV.A.9). Below, we describe each of the practices, and use our questions to determine how specific they should be at most and who should specify them. Figure 5 summarizes the results of our analysis.

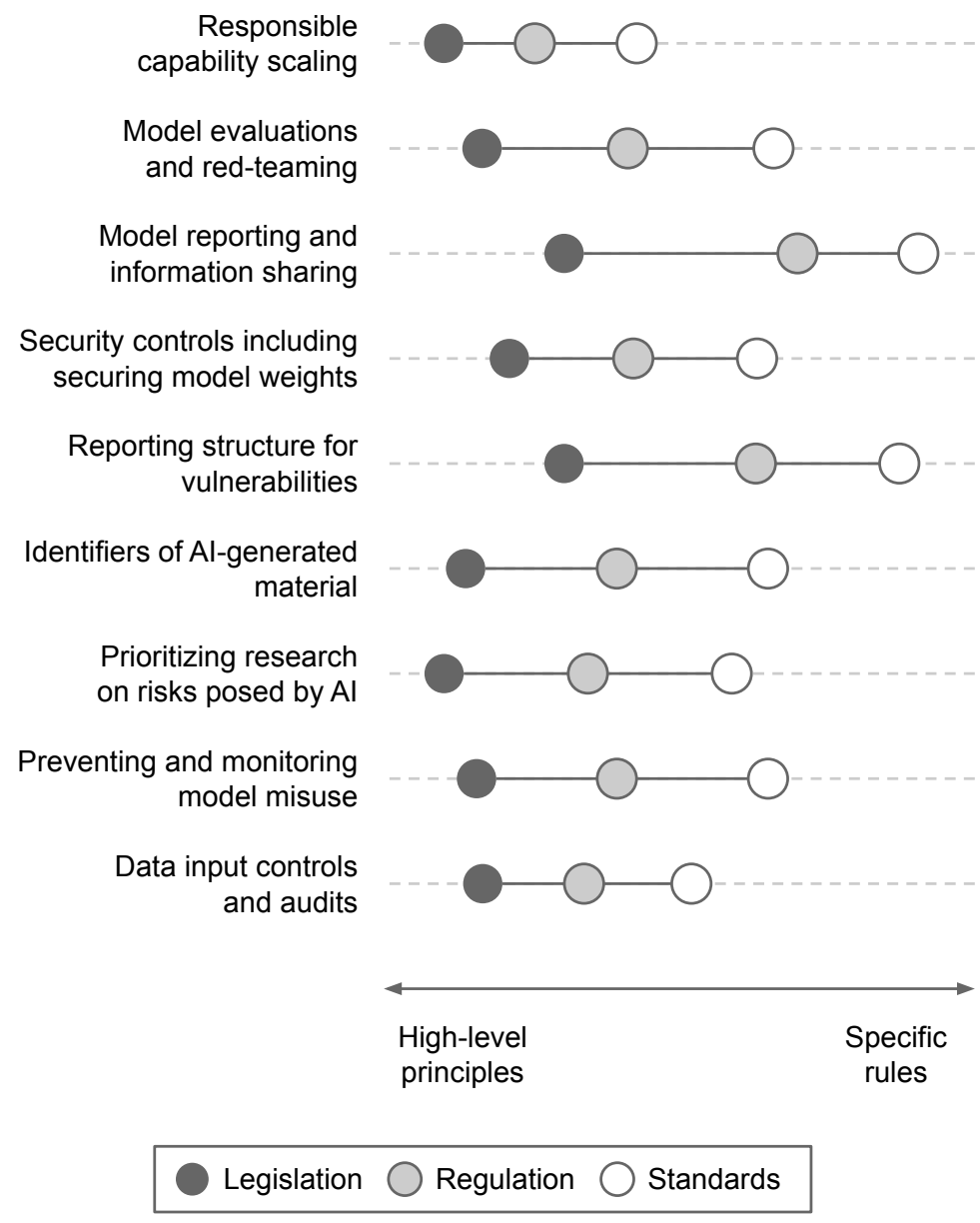

*Figure 5*: Applying our framework to the nine safety practices from DSIT's policy paper Emerging Processes for Frontier AI Safety

### *1. Responsible capability scaling*

Frontier AI developers may be required to implement a responsible capability scaling policy, i.e. an internal policy intended to keep risks from the development and deployment of frontier AI systems to an acceptable level.[135] Such policies typically contain (1) a description of specific model capabilities that would require additional safety measures ("capability thresholds"), (2) a commitment to evaluate models for dangerous capabilities, (3) a protocol that specifies what safety measures developers would implement if they discover dangerous model capabilities, and (4) a commitment to pause the development and deployment process if they are not able to implement the pre-specified safety measures. Some AI companies have already implemented such policies, including Anthropic's Responsible Scaling Policy (RSP),[136] OpenAI's Preparedness Framework,[137] and Google DeepMind's Frontier Safety Framework.[138] At the AI Seoul Summit 2024, 16 AI companies committed to implement similar policies.[139]

A requirement to implement a responsible capability scaling policy should likely remain rather abstract for now. This is because (1) many of the risks the policy seeks to address are poorly understood and rapidly evolving,[140] (2) attempts to set capability thresholds are nascent,[141] (3) best practices for dangerous model evaluations do not yet exist,[142] and (4) experts disagree widely on what safety measures would be adequate for what dangerous model capabilities. An abstract version of the requirement might state that developers must implement and maintain a responsible capability scaling policy, along with a high-level description of the policy's purpose and key components. A more specific version might set specific capability thresholds, prescribe when to run what model evaluations, etc.

[135] This type of policy was initially developed by METR, especially Paul Christiano, see METR, 'Responsible Scaling Policies (RSPs)' (2023) <https://metr.org/blog/2023-09-26-rsp> accessed 1 July 2024; see also P Christiano, 'Thoughts on Responsible Scaling Policies and Regulation' (*AI Alignment Forum*, 25 October 2023) <https://www.alignmentforum.org/posts/dxgEaDrEBkkE96CXr/thoughts-on-responsible-scaling-policies-and-regulation> accessed 1 July 2024.

[136] Anthropic, 'Responsible Scaling Policy' (2023) <https://www.anthropic.com/news/anthropics-responsible-scaling-policy> accessed 1 July 2024; see also Anthropic, 'Reflections on our Responsible Scaling Policy' (2024) <https://www.anthropic.com/news/reflections-on-our-responsible-scaling-policy> accessed 1 July 2024.

[137] OpenAI, 'Preparedness' (2023) <https://openai.com/safety/preparedness> accessed 1 July 2024.

[138] Google DeepMind, 'Introducing the Frontier Safety Framework' (2024) <https://deepmind.google/discover/blog/introducing-the-frontier-safety-framework> accessed 1 July 2024.

[139] DSIT, 'Frontier AI Safety Commitments, AI Seoul Summit 2024' (2024) <https://www.gov.uk/government/publications/frontier-ai-safety-commitments-ai-seoul-summit-2024> accessed 1 July 2024.

[140] See Section II.A.1.

[141] For more information on different types of thresholds, see L Koessler and others, 'Risk Thresholds for Frontier AI' (arXiv, 2024) <https://arxiv.org/abs/2406.14713>; M Pistillo and others, 'The Role of Compute Thresholds for AI Governance' (forthcoming); L Heim and L Koessler, 'Training Compute Thresholds: Features and Functions in AI Regulation' (forthcoming).

[142] See Section IV.A.2.

Neither legislators, regulators, or standard-setting bodies are well-placed to specify the requirement. They currently have insufficient expertise, which mostly sits inside AI companies. Companies also have access to proprietary information (e.g. about model capabilities and the effectiveness of safety measures), which makes it challenging for external parties to develop detailed requirements. Furthermore, the requirement may need to be updated more quickly than these bodies typically allow. Taken together, these considerations suggest that, instead of requiring developers to implement a specific version of the policy, regulators should initially focus on overseeing the policies developed by AI companies. The UK Government seems to be moving in this direction. Ahead of the UK AI Safety Summit, they requested frontier AI developers to outline their current responsible capability scaling policies.[143] In our view, legislation could be at least as specific as the Frontier AI Safety Commitments that a number of companies made in Seoul.[144]

*2. Model evaluations and red-teaming*

Frontier AI developers may be required to conduct model evaluations[145] and red-teaming exercises, which are methods for empirically testing the properties of a system.[146] They can be used to gather evidence about a model's

---

[143] UK Government, 'Policy Updates' (2023) <https://www.aisafetysummit.gov.uk/policy-updates> accessed 1 July 2024.

[144] DSIT, 'Frontier AI Safety Commitments, AI Seoul Summit 2024' (2024) <https://www.gov.uk/government/publications/frontier-ai-safety-commitments-ai-seoul-summit-2024> accessed 1 July 2024.

[145] For more information on model evaluations for dangerous capabilities, see T Shevlane and others, 'Model Evaluation for Extreme Risks' (arXiv, 2023) <http://arxiv.org/abs/2305.15324>; M Kinniment and others, 'Evaluating Language-Model Agents on Realistic Autonomous Tasks' (arXiv, 2023) <https://arxiv.org/abs/2312.11671>; M Phuong and others, 'Evaluating Frontier Models for Dangerous Capabilities' (arXiv, 2024) <https://arxiv.org/abs/2403.13793>; R Laine and others, 'Me, Myself, and AI: The Situational Awareness Dataset (SAD) for LLMs' (arXiv, 2024) <https://arxiv.org/abs/2407.04694>. For more information on model evaluations more generally, see M Chen and others, 'Evaluating Large Language Models Trained on Code' (arXiv, 2021) <http://arxiv.org/abs/2107.03374>; E Perez and others, 'Discovering Language Model Behaviors With Model-Written Evaluations' (arXiv, 2022) <http://arxiv.org/abs/2212.09251>; P Liang and others, 'Holistic Evaluation of Language Models' (arXiv, 2022) <http://arxiv.org/abs/2211.09110>; S Gehrmann and others, 'Repairing the Cracked Foundation: A Survey of Obstacles in Evaluation Practices for Generated Text' (arXiv, 2022) <http://arxiv.org/abs/2202.06935>.

[146] In a survey of AI safety and governance experts ($N$ = 51), 98% of respondents somewhat or strongly agreed with the statements: 'AGI labs should run evaluations to assess their models' dangerous capabilities (e.g. misuse potential, ability to manipulate, and power-

capabilities, including potentially dangerous capabilities, which can emerge unintentionally and unpredictably during training.[147] As mentioned above, model evaluations play a key role in responsible capability scaling policies.[148]

Model evaluation and red-teaming requirements should likely be rather abstract for now. Although they are fairly popular,[149] they have only recently been applied to risks from frontier AI systems. The underlying science is nascent and rapidly evolving, currently more art than science. Best practices do not yet exist. A particularly challenging aspect is elicitation, i.e. the process of extracting a model's latent capabilities through techniques such as tool use, prompting, scaffolding, and fine-tuning. Elicitation is crucial for reliable evaluations, as it can significantly improve observed model capabilities. Against this background, overly specific requirements may quickly become obsolete. The most abstract version of the requirement might state that companies must use adequate methods for empirically testing the properties of their systems. A slightly more specific version could explicitly mention model evaluations and red-teaming, along with a high-level description of their purpose and the contours of a testing regime (e.g. what types of capabilities to

---

seeking behavior)' and 'AGI labs should commission external red teams before deploying powerful models', see J Schuett and others, 'Towards Best Practices in Agi Safety and Governance: A Survey of Expert Opinion' (arXiv, 2023) <https://arxiv.org/abs/2305.07153>.

[147] D Ganguli and others, 'Predictability and Surprise in Large Generative Models' (ACM Conference on Fairness, Accountability, and Transparency, 2022) <https://doi.org/10.1145/3531146.3533229>; J Wei and others, 'Emergent Abilities of Large Language Models' (arXiv, 2022) <https://arxiv.org/abs/2206.07682>. But note that it has been claimed that emergent capabilities may not be a fundamental property of scaling AI models, see R Schaeffer, B Miranda and S Koyejo, 'Are Emergent Abilities of Large Language Models a Mirage?' (Conference on Neural Information Processing Systems, 2023) <https://arxiv.org/abs/2304.15004>.

[148] See Section IV.A.1.

[149] For more information on red-teaming, see D Ganguli and others, 'Red Teaming Language Models to Reduce Harms: Methods, Scaling Behaviors, and Lessons Learned' (arXiv, 2022) <https://arxiv.org/abs/2209.07858>; E Perez and others, 'Red Teaming Language Models with Language Models' (arXiv, 2022) <https://arxiv.org/abs/2202.03286>; B Radharapu and others, 'AART: AI-Assisted Red-Teaming with Diverse Data Generation for New LLM-Powered Applications' (arXiv, 2023) <https://arxiv.org/abs/2311.08592>; CA Mouton, C Lucas and E Guest, 'The Operational Risks of AI in Large-Scale Biological Attacks: A Red-Team Approach' (*RAND*, 2023) <https://doi.org/10.7249/RRA2977-1>; Anthropic, 'Frontier Threats Red Teaming for AI Safety' (2023) <https://www.anthropic.com/news/frontier-threats-red-teaming-for-ai-safety> accessed 1 July 2024; Anthropic, 'Challenges in Red Teaming AI Systems' (2024) <https://www.anthropic.com/news/challenges-in-red-teaming-ai-systems> accessed 1 July 2024; M Samvelyan, 'Rainbow Teaming: Open-Ended Generation of Diverse Adversarial Prompts' (arXiv, 2024) <https://arxiv.org/abs/2402.16822>; L Weidinger and others, 'STAR: SocioTechnical Approach to Red Teaming Language Models' (arXiv, 2024) <https://arxiv.org/abs/2406.11757>.

evaluate, that independent third parties should be involved, and that results should inform deployment decisions). The most specific version would prescribe a detailed testing regime.

This abstract approach should extend to legislative, regulatory, and standard-setting levels, including NIST's task from the Executive Order on Safe, Secure, and Trustworthy AI to create guidance and benchmarks for evaluating AI capabilities.[150] This is because AI companies have significant advantages in conducting evaluations, particularly elicitation, making it difficult for government actors to specify evaluation methods. Companies possess superior access to top technical talent and useful proprietary information, including insider knowledge of which techniques work best for their specific models. Additionally, optimal elicitation approaches can differ, sometimes significantly, between AI models, meaning employees of a particular developer are uniquely positioned to effectively elicit capabilities from their own models.

### *3. Model-reporting and information-sharing*

Frontier AI developers may also be required to disclose certain information to relevant stakeholders (e.g. government, the public, or civil society).[151] This can include information about their systems (e.g. via system cards,[152] model cards,[153] or data sheets[154]), about their safety measures (e.g. via safety cases[155]), or about safety and security incidents (e.g. via incident databases[156] or special information-sharing regimes with government[157]).

---

[150] White House, 'Safe, Secure, and Trustworthy Development and Use of Artificial Intelligence' (2023) Executive Order 14110 <https://www.federalregister.gov/documents/2023/11/01/2023-24283/safe-secure-and-trustworthy-development-and-use-of-artificial-intelligence> accessed 1 July 2024.

[151] N Kolt, 'Responsible Reporting for Frontier AI Development' (arXiv, 2024) <https://arxiv.org/abs/2404.02675>.

[152] N Green and others, 'System Cards: A New Resource for Understanding How AI Systems Work' (*Meta AI*, 23 February 2022) <https://ai.meta.com/blog/system-cards-a-new-resource-for-understanding-how-ai-systems-work> accessed 1 July 2024.

[153] M Mitchell and others, 'Model Cards for Model Reporting' (Conference on Fairness, Accountability, and Transparency, 2019) <https://doi.org/10.1145/3287560.3287596>.

[154] T Gebru and others, 'Datasheets for Datasets' (2021) 64 Communications of the ACM 86 <https://doi.org/10.1145/3458723>; E Bender and B Friedman, 'Data Statements for Natural Language Processing: Toward Mitigating System Bias and Enabling Better Science' (2018) 6 Transactions of the Association for Computational Linguistics 587 <https://doi.org/10.1162/tacl_a_00041>.

[155] J Clymer and others, 'Safety Cases: How to Justify the Safety of Advanced AI Systems' (arXiv, 2024) <https://arxiv.org/abs/2403.10462>; A Wasil and others, 'Affirmative Safety: An Approach to Risk Management for Advanced AI' (SSRN, 2024) <https://doi.org/10.2139/ssrn.4806274>; MD Buhl and others, 'Safety Cases for Frontier AI' (forthcoming).

Model-reporting and information-sharing requirements should likely be quite specific. The relevant categories of information for governments to request are relatively clear, and certain practices for public reporting, such as model cards, are already widely adopted. Moreover, companies' incentives may significantly diverge from the public good due to concerns about intellectual property, favoring specificity when it comes to these requirements.

Regulators and standard-setting bodies are capable of specifying model-reporting and information-sharing requirements. In fact, US regulators implementing the reporting requirements in the Executive Order on Safe, Secure, and Trustworthy AI are expected to be quite specific. Moreover, regulators may not need to delegate to standard-setting bodies, as collecting information is a common regulatory practice, and they are less likely to be influenced by industry interests compared to standard-setting bodies. Having said that, regulators would still benefit from input provided by civil society and companies when determining the specifics of their inquiries.

*4. Security controls including securing model weights*

Frontier AI regulation may require developers to implement certain security controls, i.e. cyber, personnel, and physical security measures to protect AI systems from being stolen or leaked. Some consider model weights, the learnable parameters of models, especially important to secure, as they can enable malicious actors to bypass protections and exploit the model's capabilities at a fraction of the training cost.[158] Existing cybersecurity practices, such as the NIST Secure Software Development Framework (SSDF),[159] National Cyber Security Centre (NCSC) guidance,[160] and National

[156] S McGregor, 'Preventing Repeated Real World AI Failures by Cataloguing Incidents: The AI Incident Database' (AAAI Conference on Artificial Intelligence, 2021) <https://doi.org/10.1609/aaai.v35i17.17817>; OECD, 'AI Incidents Monitor' (2024) <https://oecd.ai/en/incidents> accessed 1 July 2024.

[157] N Kolt and others, 'Responsible Reporting for Frontier AI Development' (arXiv, 2024) <https://arxiv.org/abs/2404.02675>; N Mulani and J Whittlestone, 'Proposing a Foundation Model Information-Sharing Regime for the UK' (*Centre for the Governance of AI*, 16 June 2023) <https://www.governance.ai/post/proposing-a-foundation-model-information-sharing-regime-for-the-uk> accessed 1 July 2024.

[158] S Nevo and others, 'Securing Artificial Intelligence Model Weights' (*RAND*, 2024) <https://doi.org/10.7249/RRA2849-1>; Anthropic, 'Frontier Model Security' (2023) <https://www.anthropic.com/news/frontier-model-security> accessed 1 July 2024.

[159] NIST, 'Secure Software Development Framework' (2024) <https://csrc.nist.gov/projects/ssdf> accessed 1 July 2024.

[160] National Cyber Security Centre, 'How to Assess and Gain Confidence in Your Supply Chain Cyber Security' (2022) <https://www.ncsc.gov.uk/collection/assess-supply-chain-cyber-security> accessed 1 July 2024.

Protective Security Authority (NSPA) guidance,[161] are relevant to AI system security, but the unique vulnerabilities of AI systems may warrant further controls.[162]

Security control requirements should likely be somewhere in the middle of the spectrum. Security controls can be extremely complex and depend on implementation details, making it difficult to set effective general requirements. Moreover, there is considerable uncertainty about the feasibility of AI-specific attack vectors and effectiveness of AI-specific mitigations, largely because the AI security field is emerging and evolving.[163] However, despite this, experts have recently been able to draft somewhat prescriptive guidelines addressing AI security.[164] This is likely because national security experts possess a wealth of (sometimes classified) knowledge about security, which can be repurposed to address AI security.

Regulators and standard-setting bodies may be able to effectively specify AI security requirements by leveraging the expertise of the national security community. This community possesses extensive security knowledge and experience in addressing sophisticated threats, including those from state actors, which may be unfamiliar to many industry participants. Incorporating this specialized knowledge will be key to creating effective security requirements for AI systems.

*5. Reporting structure for vulnerabilities*

Frontier AI developers could be required to set up a reporting structure that allows external parties to report vulnerabilities they discover in AI systems. Some of these vulnerabilities will be akin to traditional software vulnerabilities. For example, in March 2023, OpenAI had an incident where some users could temporarily view other users' chat history.[165] Other vulnerabilities will be more closely related to AI models (e.g. unanticipated jailbreaks[166] or prompt injection attacks[167]), which may persist in AI systems despite extensive pre-deployment testing.[168] One example of a vulnerability reporting structure is Google's Vulnerability Reward Program, which encourages individuals to report vulnerabilities they discover in Google's AI products in exchange for potential rewards.[169] Another reporting structure is OpenAI's Bug Bounty Program, aimed specifically at vulnerabilities in the software surrounding their AI models, rather than the models themselves.[170]

---

[161] National Protective Security Authority, 'Insider Risk' (2024) <https://www.npsa.gov.uk/insider-risk> accessed 28 March 2024.

[162] S Nevo and others, 'Securing Artificial Intelligence Model Weights' (*RAND*, 2024) <https://doi.org/10.7249/RRA2849-1>; Anthropic, 'Frontier Model Security' (2023) <https://www.anthropic.com/news/frontier-model-security> accessed 1 July 2024.

[163] S Nevo and others, 'Securing Artificial Intelligence Model Weights' (*RAND*, 2024) <https://doi.org/10.7249/RRA2849-1>.

[164] S Nevo and others, 'Securing Artificial Intelligence Model Weights' (*RAND*, 2024) <https://doi.org/10.7249/RRA2849-1>.

[165] E Kovacs, 'ChatGPT Data Breach Confirmed as Security Firm Warns of Vulnerable Component Exploitation' (*SecurityWeek*, 28 March 2023) <https://www.securityweek.com/chatgpt-data-breach-confirmed-as-security-firm-warns-of-vulnerable-component-exploitation> accessed 1 July 2024.

Requirements for reporting structures for traditional software vulnerabilities can be relatively specific, as many norms from well-established bug bounty programs can be adapted. Best practices from established programs include: (1) providing a safe harbor to individuals who report vulnerabilities, (2) clearly stating what vulnerabilities are sought and what compensation different kinds of vulnerabilities entail, and (3) allowing for responsible disclosure (where vulnerabilities can be made public within a certain number of days after having been reported to the company, thereby incentivizing speedy patching).

For AI-specific vulnerabilities (e.g. a model leaking sensitive training data in its outputs), it is less clear what appropriate reporting structures would look like. This is because it is often unclear how to patch such vulnerabilities and what the offense-defense balance of relevant knowledge is.[171] As such, public disclosure of such vulnerabilities could be harmful. However, at minimum, it seems that companies could have commitments to allow reporting of such vulnerabilities, responding to such vulnerabilities in a timely manner where possible, and provide safe harbors for vulnerability identification and red teaming – ensuring researchers can do such work without fear of reprisals.[172]

---

[166] C Anil and others, 'Many-Shot Jailbreaking' (*Anthropic*, 2024) <https://www.anthropic.com/research/many-shot-jailbreaking> accessed 1 July 2024.

[167] K Greshake and others, 'Not What You've Signed Up For: Compromising Real-World LLM-Integrated Applications with Indirect Prompt Injection' (ACM Workshop on Artificial Intelligence and Security, 2023) <https://doi.org/10.1145/3605764.3623985>.

[168] S McGregor, 'Preventing Repeated Real World AI Failures by Cataloguing Incidents: The AI Incident Database' (AAAI Conference on Artificial Intelligence, 2021) <https://doi.org/10.1609/aaai.v35i17.17817>; OECD, 'AI Incidents Monitor' (2024) <https://oecd.ai/en/incidents> accessed 1 July 2024.

[169] Google, 'Google and Alphabet Vulnerability Reward Program (VRP) Rules' <https://bughunters.google.com/about/rules/google-friends/6625378258649088/google-and-alphabet-vulnerability-reward-program-vrp-rules> accessed 1 July 2024.

[170] OpenAI, 'Announcing OpenAI's Bug Bounty Program' (2023) <https://openai.com/blog/bug-bounty-program> accessed 1 July 2024.

[171] T Shevlane and A Dafoe, 'The Offense-Defense Balance of Scientific Knowledge: Does Publishing AI Research Reduce Misuse?' (AAAI/ACM Conference on AI, Ethics, and Society, 2020) <https://doi.org/10.1145/3375627.3375815>.

[172] S Longpre and others, 'A Safe Harbor for AI Evaluation and Red Teaming' (arXiv, 2024) <https://arxiv.org/abs/2403.04893>. See also R Gruetzemacher and others, 'An International Consortium for Evaluations of Societal-Scale Risks from Advanced AI' (arXiv, 2023) <https://arxiv.org/abs/2310.14455>.

Both regulators and standard-setting bodies can specify these requirements. Regulators could do significant parts of that specification with regards to software vulnerabilities, where best practices are more established. However, with regards to AI-specific vulnerabilities, they might benefit from delegating more work to standard-setting bodies and the companies themselves.

*6. Identifiers of AI-generated material*

Developers could also be required to use identifiers that allow the detection of AI-generated content. One prominent example is watermarking, which involves embedding signatures into AI-generated content.[173] Watermarking and other identifiers can be used to flag AI-generated disinformation and other deceptive content. However, current techniques are still somewhat unreliable. For example, watermarks are not consistently robust to tampering.[174]

Requirements for identifiers of AI-generated material should plausibly be rather specific, largely because they could adapt pre-existing standards, developed in other fields facing similar challenges, which are useful for the AI context. These include the Coalition for Content Provenance and Authenticity (C2PA)[175] and NIST Digital Identity Guidelines.[176] However, new standards are necessary to address emerging AI-related issues. One example is the potential proliferation of AI agents, for which agent-specific watermarks could be used to attribute content.[177] Since best practices do not currently exist, these standards should likely be somewhat abstract.

[173] Z Wang and others, 'Data Hiding with Deep Learning: A Survey Unifying Digital Watermarking and Steganography' (arXiv, 2021) <https://arxiv.org/abs/2107.09287>; X Zhong and others, 'A Brief Yet In-Depth Survey of Deep Learning-Based Image Watermarking' (arXiv, 2023) <https://arxiv.org/abs/2308.04603>; SS Ghosal and others, 'Towards Possibilities & Impossibilities of AI-generated Text Detection: A Survey' (arXiv, 2023) <https://arxiv.org/abs/2310.15264>; J Kirchenbauer and others, 'A Watermark for Large Language Models' (International Conference on Machine Learning, 2023) <https://arxiv.org/abs/2301.10226>; X Zhao and others, 'Protecting Language Generation Models via Invisible Watermarking' (International Conference on Machine Learning, 2023) <https://arxiv.org/abs/2302.03162>; M Saberi and others, 'Robustness of AI-Image Detectors: Fundamental Limits and Practical Attacks' (arXiv, 2023) <https://arxiv.org/abs/2310.00076>; H Zhang and others, 'Watermarks in the Sand: Impossibility of Strong Watermarking for Generative Models' (arXiv, 2023) <https://arxiv.org/abs/2311.04378>; P Fernandez and others, 'The Stable Signature: Rooting Watermarks in Latent Diffusion Models' (International Conference on Computer Vision, 2023) <https://arxiv.org/abs/2303.15435>; Google DeepMind, 'Watermarking AI-Generated Text and Video With SynthID' (2024) <https://deepmind.google/discover/blog/watermarking-ai-generated-text-and-video-with-synthid> accessed 1 July 2024; A Liu and others, 'A Survey of Text Watermarking in the Era of Large Language Models' (arXiv, 2024) <https://arxiv.org/abs/2312.07913>.

[174] M Saberi and others, 'Robustness of AI-Image Detectors: Fundamental Limits and Practical Attacks' (arXiv, 2023) <https://arxiv.org/abs/2310.00076>.

[175] C2PA, <https://c2pa.org> accessed 1 July 2024.

[176] NIST, 'Digital Identity Guidelines' (2020) Special Publication 800-63-3 <https://doi.org/10.6028/NIST.SP.800-63-3>.

Standard-setting bodies are well-positioned to identify and adapt existing best practices relevant to AI. However, a more oversight-heavy, principle-based approach seems advisable to bring to bear the expertise of AI companies.

*7. Prioritizing research on risks posed by AI*

Frontier AI developers could be required to conduct research on the risks associated with frontier AI systems. This might include AI safety research in areas such as alignment,[178] interpretability,[179] and evaluation.[180] Developers could also be required to develop risk assessment and risk mitigation techniques as well as practices to enable relevant research (e.g. providing model access to external researchers and sharing research results[181]).

[177] A Chan and others, 'Harms from Increasingly Agentic Algorithmic Systems' (ACM Conference on Fairness, Accountability, and Transparency, 2023) <https://doi.org/10.1145/3593013.3594033>; A Chan and others, 'Visibility Into AI Agents' (ACM Conference on Fairness, Accountability, and Transparency, 2024) <https://doi.org/10.1145/3630106.3658948>; Y Shavit and others, 'Practices for Governing Agentic AI Systems' (*OpenAI*, 2023) <https://openai.com/index/practices-for-governing-agentic-ai-systems> accessed 1 July 2024; MK Cohen and others, 'Regulating Advanced Artificial Agents' (2024) 384 Science 36 <https://doi.org/10.1126/science.adl0625>; I Gabriel and others, 'The Ethics of Advanced AI Assistants' (arXiv, 2024) <https://arxiv.org/abs/2404.16244>; N Kolt, 'Governing AI Agents' (SSRN, 2024) <https://doi.org/10.2139/ssrn.4772956>.

---

[178] See e.g. P Christiano and others, 'Deep Reinforcement Learning from Human Preferences' (arXiv, 2017) <http://arxiv.org/abs/1706.03741>; P Christiano, B Shlegeris and D Amodei, 'https://arxiv.org/abs/1810.08575' (arXiv, 2018) <https://arxiv.org/abs/1810.08575>; J Leike and others, 'Scalable Agent Alignment via Reward Modeling: A Research Direction' (arXiv, 2018) <https://arxiv.org/abs/1811.07871>; G Irving, P Christiano and D Amodei, 'AI Safety via Debate' (arXiv, 2018) <https://arxiv.org/abs/1805.00899>; DM Ziegler and others, 'Fine-Tuning Language Models from Human Preferences' (arXiv, 2019) <http://arxiv.org/abs/1909.08593>; S Russell, *Human Compatible: Artificial Intelligence and the Problem of Control* (Penguin Random House 2019); B Christian, *The Alignment Problem: Machine Learning and Human Values* (W. W. Norton & Company 2020); I Gabriel, 'Artificial Intelligence, Values, and Alignment' (2020) 30 Minds and Machines 411 <https://doi.org/10.1007/s11023-020-09539-2>; D Hendrycks and others, 'Unsolved Problems in ML Safety' (arXiv, 2021) <https://arxiv.org/abs/2109.13916>; Z Kenton and others, 'Alignment of Language Agents' (arXiv, 2021) <https://arxiv.org/abs/2103.14659>; R Ngo, L Chan and S Mindermann, 'The Alignment Problem from a Deep Learning Perspective' (arXiv, 2022) <https://arxiv.org/abs/2209.00626>; S Bowman and others, 'Measuring Progress on Scalable Oversight for Large Language Models' (arXiv, 2022) <https://arxiv.org/abs/2211.03540>; Y Bai and others, 'Constitutional AI: Harmlessness from AI Feedback' (arXiv, 2022) <http://arxiv.org/abs/2212.08073>; C Burns and others, 'Weak-to-Strong Generalization: Eliciting Strong Capabilities With Weak Supervision' (arXiv, 2023) <https://arxiv.org/abs/2312.09390>; Z Kenton and others, 'On Scalable Oversight With Weak LLMs Judging Strong LLMs' (arXiv, 2024) <https://arxiv.org/abs/2407.04622>.

[179] See e.g. N Nanda and others, 'Progress Measures for Grokking via Mechanistic Interpretability' (arXiv, 2023) <https://arxiv.org/abs/2301.05217>; C Olah, 'Interpretability Dreams' (*Transformer Circuits Thread*, 2023) <https://transformer-circuits.pub/2023/interpretability-dreams/index.html> accessed 1 July 2024; T Bricken and others, 'Towards Monosemanticity: Decomposing Language Models With Dictionary Learning' (*Transformer Circuits Thread*, 2023) <https://transformer-circuits.pub/2023/monosemantic-features/index.html> accessed 1 July 2024; A Templeton and others, 'Scaling Monosemanticity: Extracting Interpretable Features from Claude 3 Sonnet' (*Transformer Circuits Thread*, 2024) <https://transformer-circuits.pub/2024/scaling-monosemanticity/index.html> accessed 1 July 2024.

[180] See e.g. T Shevlane and others, 'Model Evaluation for Extreme Risks' (arXiv, 2023) <http://arxiv.org/abs/2305.15324>; M Kinniment and others, 'Evaluating Language-Model Agents on Realistic Autonomous Tasks' (arXiv, 2023) <https://arxiv.org/abs/2312.11671>; M Phuong and others, 'Evaluating Frontier Models for Dangerous Capabilities' (arXiv, 2024) <https://arxiv.org/abs/2403.13793>; R Laine and others, 'Me, Myself, and AI: The Situational Awareness Dataset (SAD) for LLMs' (arXiv, 2024) <https://arxiv.org/abs/2407.04694>.

Requirements for prioritizing research on frontier AI risks should likely remain relatively abstract. While there is some agreement on promising research directions, there is little agreement on their relative importance and the best ways of conducting such research.[182] Since there is more agreement on first steps towards practices that enable relevant research, such requirements could be more specific with regards to, for instance, providing access to external researchers and third party model evaluators.[183]

Some government actors, like the emerging AI Safety Institutes, may be able to outline key research areas and perhaps also the (relative) amount of resources that developers should spend on them.[184] AI companies, academia, and civil society, who collectively possess much of the relevant expertise, should have significant autonomy to determine the most effective methods for conducting this research. Practices that enable relevant research could be specified by standard-setting bodies.

*8. Preventing and monitoring model misuse*

Developers could be required to implement measures to monitor and prevent misuse of frontier AI systems. Monitoring practices include human or AI review of usage logs and retaining logs for an appropriate duration. Prevention practices include restricting access to models, modifying them to improve safety (e.g. RLHF[185] or RLAIF[186]), blocking or modifying harmful inputs (e.g. with input filters using AI classifiers), and blocking or modifying harmful outputs (e.g. with system prompts or completion filters using AI classifiers).[187] They could also include readiness to rapidly de-deploy a model if necessary.

[181] T Shevlane and A Dafoe, 'The Offense-Defense Balance of Scientific Knowledge: Does Publishing AI Research Reduce Misuse?' (AAAI/ACM Conference on AI, Ethics, and Society, 2020) <https://doi.org/10.1145/3375627.3375815>.

[182] For one list of open challenges, see U Anwar and others, 'Foundational Challenges in Assuring Alignment and Safety of Large Language Models' (arXiv, 2024) <http://arxiv.org/abs/2404.09932>.

[183] See e.g. BS Bucknall and R Trager, 'Structured Access for Third-Party Research on Frontier AI Models: Investigating Researchers' Model Access Requirements' (*Centre for the Governance of AI*, 2023) <https://www.governance.ai/research-paper/structured-access-for-third-party-research-on-frontier-ai-models> accessed 1 July 2024.

[184] M Ziosi and others, 'AISIs' Roles in Domestic and International Governance' (forthcoming).

[185] P Christiano and others, 'Deep Reinforcement Learning From Human Preferences' (arXiv, 2017) <http://arxiv.org/abs/1706.03741>; DM Ziegler and others, 'Fine-Tuning Language Models From Human Preferences' (arXiv, 2019) <http://arxiv.org/abs/1909.08593>; N Lambert and others, 'Illustrating Reinforcement Learning from Human Feedback (RLHF)' (*Hugging Face*, 9 December 2022) <https://huggingface.co/blog/rlhf> accessed 1 July 2024.

[186] Y Bai and others, 'Constitutional AI: Harmlessness from AI Feedback' (arXiv, 2022) <http://arxiv.org/abs/2212.08073>.

These requirements should likely be quite abstract. The most effective techniques for prevention and monitoring are likely to change over time and, as model capabilities improve, additional practices will likely need to be developed. For example, mechanistic interpretability, which seeks to understand AI models' internal workings, could eventually be used to assess a model's potential for misuse and evaluate the effectiveness of mitigations.[188] Similarly, much work focused on increasing refusal of harmful model requests over the past few years has relied on RLHF,[189] but recent research suggests that interventions aimed at the model's representations of harmful behavior may be more effective.[190]

For these reasons, standard-setting bodies should likely not specify which practices AI developers must use. Instead, they could specify objectives that AI developers must meet (e.g. models achieving a certain level of robustness to jailbreaking).

---

[187] M Anderljung and J Hazell, 'Protecting Society From AI Misuse: When Are Restrictions on Capabilities Warranted?' (arXiv, 2023) <https://arxiv.org/abs/2303.09377>; B Clifford, 'Preventing AI Misuse: Current Techniques' (*Centre for the Governance of AI*, 17 December 2023) <https://www.governance.ai/post/preventing-ai-misuse-current-techniques> accessed 1 July 2024; JA Goldstein and others, 'Generative Language Models and Automated Influence Operations: Emerging Threats and Potential Mitigations' (arXiv, 2023) <https://arxiv.org/abs/2301.04246>; M Brundage and others, 'The Malicious Use of Artificial Intelligence: Forecasting, Prevention, and Mitigation' (arXiv, 2018) <https://arxiv.org/abs/1802.07228>.

[188] For more information on mechanistic interpretability, see e.g. N Nanda and others, 'Progress Measures for Grokking via Mechanistic Interpretability' (arXiv, 2023) <https://arxiv.org/abs/2301.05217>; C Olah, 'Interpretability Dreams' (*Transformer Circuits Thread*, 2023) <https://transformer-circuits.pub/2023/interpretability-dreams/index.html> accessed 1 July 2024; T Bricken and others, 'Towards Monosemanticity: Decomposing Language Models With Dictionary Learning' (*Transformer Circuits Thread*, 2023) <https://transformer-circuits.pub/2023/monosemantic-features/index.html> accessed 1 July 2024; A Templeton and others, 'Scaling Monosemanticity: Extracting Interpretable Features from Claude 3 Sonnet' (*Transformer Circuits Thread*, 2024) <https://transformer-circuits.pub/2024/scaling-monosemanticity/index.html> accessed 1 July 2024.

[189] P Christiano and others, 'Deep Reinforcement Learning From Human Preferences' (arXiv, 2017) <http://arxiv.org/abs/1706.03741>; DM Ziegler and others, 'Fine-Tuning Language Models From Human Preferences' (arXiv, 2019) <http://arxiv.org/abs/1909.08593>; N Lambert and others, 'Illustrating Reinforcement Learning from Human Feedback (RLHF)' (*Hugging Face*, 9 December 2022) <https://huggingface.co/blog/rlhf> accessed 1 July 2024.

[190] A Zou and others, 'Improving Alignment and Robustness with Circuit Breakers' (arXiv, 2024) <https://arxiv.org/abs/2406.04313>; A Templeton and others, 'Scaling Monosemanticity: Extracting Interpretable Features from Claude 3 Sonnet' (*Transformer Circuits Thread*, 2024) <https://transformer-circuits.pub/2024/scaling-monosemanticity/index.html> accessed 1 July 2024.

### *9. Data input controls and audits*

Finally, frontier AI developers could be required to implement data input controls and conduct data audits. These practices could be designed to detect and eliminate training data that likely contribute to dangerous capabilities of an AI system.[191] Audits can be conducted internally by AI companies or by external parties, and can leverage automated classification techniques to efficiently process large datasets. Problematic data, once identified, can be excluded from the training set.

These requirements should plausibly remain quite abstract for now. While it is widely accepted that training data influences model performance, and that removing problematic data can improve model safety, the full range of possible risks a dataset could pose is evolving and not well-understood. Due to this, requirements that are too specific run the risk of being ineffective or even harmful.

Standard-setting bodies could recommend the use of specific input control and data auditing practices. However, they should not be relied on to discover and detail training data problems, mainly because they may not possess the necessary expertise or have access to the training data used for frontier AI systems. For the time being, high-level principles seem more appropriate.

## *B. Proposing a regulatory approach for frontier AI*

Let us now take a step back to discuss common themes that have become visible through our analysis of specific practices, and propose an overarching regulatory approach for frontier AI. Below, we make high-level policy recommendations (Section IV.B.1) and state our underlying assumptions (Section IV.B.2).

---

[191] For the purposes of this chapter, we focus on training data that may contribute to the emergence of dangerous model capabilities. It goes without saying that datasets can be problematic in many other ways, see A Birhane, VU Prabhu and E Kahembwe, 'Multimodal Datasets: Misogyny, Pornography, and Malignant Stereotypes' (arXiv, 2021) <https://arxiv.org/abs/2110.01963>; A Birhane and VU Prabhu, 'Large Image Datasets: A Pyrrhic Win for Computer Vision?' (IEEE Winter Conference on Applications of Computer Vision, 2021) <https://doi.org/10.1109/WACV48630.2021.00158>; A Birhane and others, 'Into the LAION's Den: Investigating Hate in Multimodal Datasets' (Conference on Neural Information Processing Systems, 2023) <https://arxiv.org/abs/2311.03449>; A Birhane and others, 'The Dark Side of Dataset Scaling: Evaluating Racial Classification in Multimodal Models' (ACM Conference on Fairness, Accountability, and Transparency, 2024) <https://doi.org/10.1145/3630106.3658968>.

*1. Policy recommendations*

We recommend that policymakers should initially:

*Mandate adherence to high-level principles about the safe development and deployment of frontier AI systems*. For example, developers could be required to ensure that 'risks from frontier AI systems are reduced to an acceptable level' or that 'frontier AI systems are safe and secure'. One way in which developers could comply with such principles would be to implement high-quality responsible capability scaling policies.[192] However, as mentioned above,[193] the appropriate level of specificity may differ between individual requirements. Some requirements can already be formulated as specific rules. For example, it is already possible to specify categories of information developers should report about their models.[194] Yet, most requirements should initially remain somewhat abstract. Note that this also includes voluntary standards. Overspecification should be avoided at all levels in the hierarchy of law.

*Ensure that regulators and third parties closely oversee how frontier AI developers comply with high-level principles*. First, regulators need to have access to relevant information. To this end, a combination of information-sharing requirements and investigative powers will likely be necessary. For example, it might make sense to require developers to submit a safety case before deploying a new frontier system.[195] A safety case is a report that makes a structured argument, supported by evidence, that a system is sufficiently safe. It might even make sense for some regulators to be permanently located in the developers' offices. Second, regulators need to be able to correct company practices that violate the principles. To this end, they will likely need considerable flexibility and discretion. Third, regulatory oversight should be complemented by independent third parties. Among other things, these parties should evaluate frontier AI systems for dangerous capabilities and conduct governance audits to verify key claims that developers make (e.g. that they have followed their safety policies).[196]

---

[192] See Section IV.A.1.

[193] See Section III.A.1 and Section IV.A.

[194] See Section IV.A.3.

[195] J Clymer and others, 'Safety Cases: How to Justify the Safety of Advanced AI Systems' (arXiv, 2024) <https://arxiv.org/abs/2403.10462>; A Wasil and others, 'Affirmative Safety: An Approach to Risk Management for Advanced AI' (SSRN, 2024) <https://doi.org/10.2139/ssrn.4806274>; MD Buhl and others, 'A Safety Case For Frontier AI' (forthcoming).

[196] For more information on verifying claims about responsible AI development, see M Brundage and others, 'Toward Trustworthy AI Development: Mechanisms for Supporting Verifiable Claims' (arXiv, 2020) <https://arxiv.org/abs/2004.07213>; S Avin and others, 'Filling Gaps in Trustworthy Development of AI' (2021) 374 Science 1327 <https://doi.org/10.1126/science.abi7176>; J Mökander and others, 'Auditing Large

*Urgently build up regulatory capacity*. Regulators need to be able to assess whether the different ways in which developers seek to comply with high-level principles are adequate and correct them if necessary. For example, regulators need to have the expertise to assess whether a safety case sufficiently demonstrates that a system is safe enough[197] or whether a responsible capability scaling policy is of sufficient quality.[198] Eventually, they should be able to make such assessments internally, without relying on external actors for help (though it might still make sense to involve third parties).

As the facts change, so too will the best course of action. Over time, as best practices emerge and regulators build up capacity, the approach should likely become more rule-based. We also wish to emphasize that not all AI companies should be subject to this oversight-heavy regulatory regime. Instead, the regime should focus on companies that develop frontier AI systems.[199] These systems will likely pose the largest and least understood risks,[200] while also being adopted widely. For systems that pose smaller and better understood risks (i.e. the vast majority of AI systems), a less onerous approach is warranted.

*2. Assumptions*

Our recommendations are based on the following assumptions:

*Risks from frontier AI systems are poorly understood and rapidly evolving*. This favors a regulatory approach that (1) is adaptable to changes in the risk landscape, (2) requires developers to implement safety measures based on the results of risk assessments (e.g. model evaluations), and (3) grants regulators substantial discretion.

*Many safety practices are still nascent*. It is often unclear what specific behavior would best advance the regulatory objective of keeping risk to an acceptable level. For example, although model evaluations for dangerous

---

Language Models: A Three-Layered Approach' (2023) AI and Ethics <https://doi.org/10.1007/s43681-023-00289-2>.

[197] J Clymer and others, 'Safety Cases: How to Justify the Safety of Advanced AI Systems' (arXiv, 2024) <https://arxiv.org/abs/2403.10462>; A Wasil and others, 'Affirmative Safety: An Approach to Risk Management for Advanced AI' (SSRN, 2024) <https://doi.org/10.2139/ssrn.4806274>; MD Buhl and others, 'A Safety Case For Frontier AI' (forthcoming).

[198] Anthropic, 'Responsible Scaling Policy' (2023) <https://www.anthropic.com/news/anthropics-responsible-scaling-policy> accessed 1 July 2024; OpenAI, 'Preparedness' (2023) <https://openai.com/safety/preparedness> accessed 1 July 2024; Google DeepMind, 'Introducing the Frontier Safety Framework' (2024) <https://deepmind.google/discover/blog/introducing-the-frontier-safety-framework> accessed 1 July 2024.

[199] See Section II.A.3.

[200] See Section II.A.1.

capabilities play a key role in companies' safety policies,[201] best practices do not yet exist and we are still far from a rigorous science of evaluations.[202] It seems unlikely that prematurely prescribing specific safety practices will reduce risks to an acceptable level, without also curtailing positive uses of the technology.

*Frontier AI developers are best placed to innovate on safety practices*. They employ many of the world's leading AI safety researchers and have more resources than other research institutions. For example, they can provide researchers with unrestricted access to models and pay much higher salaries. Historically, many safety practices have been developed by companies (e.g. OpenAI developed RLHF[203]). A principle-based approach could incentivize companies to further invest in refining existing and developing new, more efficient, safety practices (because the approach would encourage them to decide on their own how to best comply with high-level principles). In contrast, a rules-based approach would disincentivize such investments (because developers would be required to implement prespecified safety measures).

*The incentives of frontier AI developers are not aligned with the public interest*.[204] As mentioned above,[205] market incentives will likely push developers to underinvest in safety,[206] which would be particularly concerning

---

[201] E.g. Anthropic, 'Responsible Scaling Policy' (2023) <https://www.anthropic.com/news/anthropics-responsible-scaling-policy> accessed 1 July 2024; OpenAI, 'Preparedness' (2023) <https://openai.com/safety/preparedness> accessed 1 July 2024; Google DeepMind, 'Introducing the Frontier Safety Framework' (2024) <https://deepmind.google/discover/blog/introducing-the-frontier-safety-framework> accessed 1 July 2024.

[202] M Phuong and others, 'Evaluating Frontier Models for Dangerous Capabilities' (arXiv, 2024) <https://arxiv.org/abs/2403.13793>; Apollo Research, 'We need a Science of Evals' (2024) <https://www.apolloresearch.ai/blog/we-need-a-science-of-evals> accessed 1 July 2024.

[203] P Christiano and others, 'Deep Reinforcement Learning From Human Preferences' (arXiv, 2017) <http://arxiv.org/abs/1706.03741>; DM Ziegler and others, 'Fine-Tuning Language Models From Human Preferences' (arXiv, 2019) <http://arxiv.org/abs/1909.08593>.

[204] See P Cihon and others, 'Corporate Governance of Artificial Intelligence in the Public Interest' (2021) 12 Information 275 <https://doi.org/10.3390/info12070275>; H Toner and T McCauley, 'AI Firms Mustn't Govern Themselves, Say Ex-Members of OpenAI's Board' (*The Economist*, 26 May 2024) <https://www.economist.com/by-invitation/2024/05/26/ai-firms-mustnt-govern-themselves-say-ex-members-of-openais-board> accessed 1 July 2024.

[205] See Section II.A.2.

[206] For example, if companies think that there is a 'winner-takes-all' mechanism, they might be willing to sacrifice safety in order to 'win the race', see S Armstrong, N Bostrom and C Shulman, 'Racing to the Precipice: A Model of Artificial Intelligence Development' (2016) 31 AI & Society 201 <https://doi.org/10.1007/s00146-015-0590-y>; A Askell, M Brundage and G Hadfield, 'The Role of Cooperation in Responsible AI Development' (arXiv, 2019) <https://arxiv.org/abs/1907.04534>; W Naudé and N Dimitri, 'The Race for

if frontier AI systems produce large negative externalities.[207] This misalignment of incentives suggests that frontier AI developers need to be closely overseen. Further, it is unclear whether the misalignment of incentives favors a rules-based approach – because companies cannot be trusted to decide on their own how to best comply with high-level principles – or a principle-based approach – because companies will adopt a box-ticking attitude and try to 'game the rules'.

*Regulators have limited expertise and access to information*. Much of the relevant expertise still sits in private industry (though there are some exceptions). This asymmetry favors a regulatory approach that shifts the burden to identify and develop adequate safety practices to frontier AI developers, while allowing regulators to extract expertise and information from companies. To this end, a collaborative relationship between regulators and frontier AI developers will be vital. At the same time, regulators need to be aware of industry capture concerns. They also need to build up in-house expertise.

*Regulators are better placed to assess the adequacy of company practices than to set detailed rules*. They already have or will soon have some of the expertise necessary to assess the adequacy of safety practices (e.g. whether a safety case sufficiently demonstrates that a system is safe enough[208] or whether responsible capability scaling policies are of sufficient quality[209]). In areas where they do not yet have the necessary inhouse expertise, they can ask independent domain experts for help. In any case, it is significantly easier to assess the adequacy of safety practices than it is to specify those practices. This suggests that a principle-based approach would be more feasible and preferable over a rule-based approach.

*Frontier AI developers can handle regulatory uncertainty*. Currently, there are only a handful of companies that develop frontier AI systems (e.g. OpenAI, Google DeepMind, and Anthropic). These companies have millions of customers and market caps in the billions, allowing them to handle higher

---

an Artificial General Intelligence: Implications for Public Policy' (2020) 35 AI & Society 367 <https://doi.org/10.1007/s00146-019-00887-x>.

[207] See Section II.A.1.

[208] J Clymer and others, 'Safety Cases: How to Justify the Safety of Advanced AI Systems' (arXiv, 2024) <https://arxiv.org/abs/2403.10462>; A Wasil and others, 'Affirmative Safety: An Approach to Risk Management for Advanced AI' (SSRN, 2024) <https://doi.org/10.2139/ssrn.4806274>; MD Buhl and others, 'A Safety Case For Frontier AI' (forthcoming).

[209] Anthropic, 'Responsible Scaling Policy' (2023) <https://www.anthropic.com/news/anthropics-responsible-scaling-policy> accessed 1 July 2024; OpenAI, 'Preparedness' (2023) <https://openai.com/safety/preparedness> accessed 1 July 2024; Google DeepMind, 'Introducing the Frontier Safety Framework' (2024) <https://deepmind.google/discover/blog/introducing-the-frontier-safety-framework> accessed 1 July 2024.

degrees of regulatory uncertainty, which would be introduced by principle-based regulation.

Should one or more of these assumptions be contested, a more rules-heavy approach might be preferable. Under such an approach, policymakers would attempt to specify rules that, if followed, would keep risks from frontier AI systems to an acceptable level. For example, they could specify what training procedures are allowed, enumerate specific capabilities that models are not allowed to have, or even set limits on the amount of compute that developers are allowed to use to train a model.[210] Nuclear power regulation in the US largely followed a similar pattern. The regime started with concrete rules developed by in-house experts. These rules were later modified in light of a more principle-based approach.[211] This approach might be preferable if (1) adequate safety measures exist and policymakers can formulate corresponding requirements, (2) actors other than frontier AI developers can comparably innovate on safety measures (e.g. academia), and (3) providing legal certainty is a key policy objective. All things considered, we do not believe that this is currently the case. However, it might change in the future and, at least for some measures, it is already possible to formulate specific rules.

## V. Conclusion

This chapter has made two main contributions. First, it has proposed a new framework that policymakers can use to decide how specific different requirements should be at the level of legislation, regulation, and voluntary standards. This framework is of practical relevance for policymakers, but it is also a contribution to the literature on regulatory approaches. Second, the chapter has contributed to the debate about how to regulate frontier AI systems. Based on the proposed framework, it has suggested how specific nine potential requirements should be at the different levels in the hierarchy of law. It has also made high-level policy recommendations, namely that policymakers should initially mandate adherence to high-level principles (e.g. 'risks from frontier AI systems should be reduced to an acceptable level'), while giving regulators considerable oversight and discretion over how regulatees implement these principles, and gradually moving the regime in a more rule-based direction.

[210] See Future of Life Institute, 'Pause Giant AI Experiments: An Open Letter' (2023) <https://futureoflife.org/open-letter/pause-giant-ai-experiments> accessed 1 July 2024.

[211] TR Wellock, *Safe Enough? A History of Nuclear Power and Accident Risk* (University of California Press 2021); NRC, 'History of the NRC's Risk-Informed Regulatory Programs' (2021) <https://www.nrc.gov/about-nrc/regulatory/risk-informed/history.html> accessed 1 July 2024.

Yet, we were only able to scratch the surface. Future work should focus on specific jurisdictions to account for the various nuances and pitfalls of the respective legal system. For example, our analysis did not account for the complex hierarchy of EU law,[212] which also includes common specifications[213] and guidelines by national competent authorities.[214] We also encourage more work fleshing out the specific requirements. Although DSIT's policy paper Emerging Processes for Frontier AI Safety[215] is a valuable resource, it was not meant to provide the basis for regulation. More generally, the academic debate about frontier AI regulation has just begun. There are numerous open questions, ranging from defining the regulatory target over effective enforcement measures to new models of supervision.

Developing a regulatory regime for frontier AI is extremely challenging. We hope that this chapter provides some of the necessary theoretical foundations and practical guidance to rise to this challenge.

## Acknowledgements

We are grateful for valuable feedback and suggestions from Peter Wills, Mackenzie Arnold, Suzanne Van Arsdale, Bilva Chandra, Matthijs Maas, José Villalobos, Marie Buhl, Patrick Levermore, Noam Kolt, Lennart Heim, Tim Fist, Gaurav Sett, Alan Chan, Christoph Winter, Yogev Bar-On, Andrew Trask, Haydn Belfield, Sella Nevo, Amelie Berz, and Philipp Hacker. All remaining errors are our own.

---

[212] For more information on the hierarchy of EU law, see D Curtin and T Manucharyan, 'Legal Acts and Hierarchy of Norms in EU Law' in D Chalmers and A Arnull (eds), *The Oxford Handbook of European Union Law* (Oxford University Press 2015) <https://doi.org/10.1093/oxfordhb/9780199672646.013.19>.

[213] The term 'common specification' is defined in Article 3, point 28 of the EU AI Act.

[214] See Article 59 of the EU AI Act.

[215] DSIT, 'Emerging Processes for Frontier AI Safety' (2023) <https://www.gov.uk/government/publications/emerging-processes-for-frontier-ai-safety> accessed 1 July 2024.